\documentclass[10pt,a4paper,superscriptaddress,aps,prd,showkeys,showpacs,nofootinbib,reprint,preprintnumbers]{revtex4-2}
\usepackage[utf8]{inputenc}
\usepackage{verbatim}
\usepackage{amsmath,amsfonts,amssymb}
\usepackage[breaklinks,colorlinks]{hyperref}
\usepackage{natbib}
\usepackage{tikz}
\usepackage{float}
\usepackage{graphicx}
\usepackage{adjustbox}
\usepackage{xcolor}
\usepackage{soul}
\usepackage{cancel}

\newcommand{\ie}{{\emph{i.e.~}}}
\newcommand{\eg}{{\emph{e.g.~}}}


\usepackage{tabularx}
\usepackage{amsbsy}
\usepackage{braket}
\hypersetup{%
,urlcolor=blue
,citecolor=blue
,linkcolor=blue
}

\def\aap{Astronomy and Astrophysics}

\def\jcap{Journal of Cosmology and Astro-Particle Physics}

\begin{document}


\title{Generalized Ray Tracing for Axions in Astrophysical Plasmas}

\author{J.~I.~McDonald}
\email[]{jamie.mcdonald@uclouvain.be}
\affiliation{Centre for Cosmology, Particle Physics and Phenomenology,
Universit\'{e} Catholique de Louvain,
Chemin du cyclotron 2,
Louvain-la-Neuve B-1348, Belgium}

\preprint{IRMP-CP3-23-49}

\author{S.~J.~Witte}
\email{samuel.witte@physics.ox.ac.uk}
\affiliation{Rudolf Peierls Centre for Theoretical Physics, University of Oxford, Parks Road, Oxford OX1 3PU, UK}
\affiliation{Departament de F\'{i}sica Qu\`{a}ntica i Astrof\'{i}sica and Institut de Ciencies del Cosmos (ICCUB) ,
Universitat de Barcelona, Diagonal 647, E-08028 Barcelona, Spain}
\affiliation{Gravitation Astroparticle Physics Amsterdam (GRAPPA), Institute for Theoretical Physics Amsterdam and Delta Institute for Theoretical Physics, University of Amsterdam, Science Park 904, 1098 XH Amsterdam, The Netherlands}

\label{firstpage}

\date{\today}

\begin{abstract}
Ray tracing plays a vital role in black hole imaging,  modeling the emission mechanisms of pulsars, and deriving signatures from physics beyond the Standard Model. In this work we focus on one specific application of ray tracing, 
namely, predicting radio signals
generated from the resonant conversion of axion dark matter in the strongly magnetized plasma surrounding neutron stars. The production and propagation of low-energy photons in these environments are sensitive to both the anisotropic response of the background plasma and curved spacetime; here, we employ a fully covariant framework capable of treating both effects. We implement this both via forward and backward ray tracing. In forward ray tracing, photons are sampled at the point of emission and propagated to infinity, whilst in the backward-tracing approach, photons are traced backwards from an image plane to the point of production. We explore various approximations adopted in prior work, quantifying the importance of gravity, plasma anisotropy, the neutron star mass and radius, and imposing the proper kinematic matching of the resonance. Finally, using a more realistic model for the charge distribution of magnetar magnetospheres, we revisit the sensitivity of current and future radio and sub-mm telescopes to spectral lines emanating from the Galactic Center Magnetar, showing such observations may extend sensitivity to axion masses $m_a \sim \mathcal{O}({\rm few}) \times 10^{-3}$ eV, potentially even probing parameter space of the QCD axion.
\end{abstract}



\maketitle




\section{Introduction}

 Astronomy is the art of inferring details about astrophysical environments through indirect measurements on the messengers they emit, be they photons, gravitational waves or neutrinos. A basic problem in astronomy is therefore to model the production and propagation of these messengers from their point of emission to the moment of detection at the observatory. Taking the case of electromagnetic signals, this entails calculating the photon production mechanism and the subsequent evolution of photons as they pass through astrophysical media. In general this requires tracking, amongst other things, their intensity, frequency, polarisation and refraction. In turn these features affect the power, directional dependence, time-variation  and spectral morphology of the signal.
 
Geometric ray tracing is a method that has been developed to accurately trace the propagation and properties of evolving wavefronts moving in inhomogeneous media or regimes of strong gravity. This technique has been successfully applied to a wide variety of different problems in astronomy and astrophysics, including, \eg,  the reconstruction of images of black holes (see \eg~\cite{Bambi:2012tg,EventHorizonTelescope:2019ggy,EventHorizonTelescope:2019pgp,Vincent:2020dij,Cardenas-Avendano:2022csp, Hu:2020usx}), the reconstruction of light curves from neutron stars (see \eg~\cite{bhattacharyya2005constraints,Cadeau:2006dc,Leahy:2011ys,psaltis2014pulse,psaltis2014prospects,Vincent:2017emv,bogdanov2019constraining}), the treatment of non-linear scattering processes of low energy photons escaping the ionosphere of the sun (see \eg~\cite{riddle1974observation,robinson1983scattering,sastry2009polarization}), and in understanding extended theories of gravity~\cite{Cropp:2013sea,Broderick:2013rlq,Mizuno:2018lxz} or particle physics~\cite{Cunha:2016bpi,Leroy:2019ghm,Witte:2021arp,Battye:2021xvt,An:2023wij}. In recent years, ray tracing has also emerged as a fundamental tool in the search for one of the most well-motivated candidates of new fundamental physics, axions~\cite{Leroy:2019ghm,Witte:2021arp,Battye:2021xvt,Noordhuis:2022ljw}.

Axions were originally introduced many decades ago to address one of the major outstanding problems in high-energy theory, the so-called Strong CP Problem~\cite{Peccei:1977hh, PQ2, WeinbergAxion, WilczekAxion}; this is effectively the question of why QCD seems to conserve charge-parity symmetry, or equivalently, why the electric dipole moment of the neutron appears to be so unnaturally small. Today, the term axion is typically used, however, to refer to the broader class of light pseudoscalars, regardless of whether they solve the strong CP problem; such particles are nevertheless well-motivated candidates for new fundamental physics, as they generically arise in  well-motivated high energy theories such as String Theory~\cite{Arvanitaki:2009fg, Witten:1984dg, Cicoli:2012sz, Conlon:2006tq, Svrcek:2006yi}.

One of the recent and more compelling proposals to indirectly search for axions in astrophysical environments involves looking for radio photons generated from axion-photon mixing in the magnetospheres of neutron stars, where the large magnetic fields and ambient plasma serve to resonantly amplify the mixing process. The presence of axions in these environments can give rise to a number of distinctive signatures at radio energies, including narrow spectral lines from the conversion of axion dark matter~\cite{Pshirkov:2007st,Hook:2018iia, Huang:2018lxq, Leroy:2019ghm, Safdi:2018oeu, Battye:2019aco, Buckley:2020fmh, Foster:2020pgt, Edwards:2020afl, Prabhu:2020yif, Foster:2022fxn, Witte:2021arp,millar2021axionphotonUPDATED, Bai:2021nrs,Nurmi:2021xds,Battye:2021yue,Witte:2022cjj,Battye:2023oac}, and large broadband emission generated from axions locally sourced in the magnetosphere itself~\cite{Prabhu:2021zve,Noordhuis:2022ljw,Noordhuis:2023wid}. In recent years, ray tracing has played an increasingly prominent role in understanding of the inhomogeneity, spectrum, and temporal evolution of these radio signals, and has proven to be a fundamental tool in accurately interpreting radio searches for axions~\cite{Witte:2021arp,Foster:2022fxn,Witte:2022cjj,Noordhuis:2022ljw,Battye:2023oac}.

\begin{figure*}
	\centering
  \includegraphics[width=0.45\textwidth]{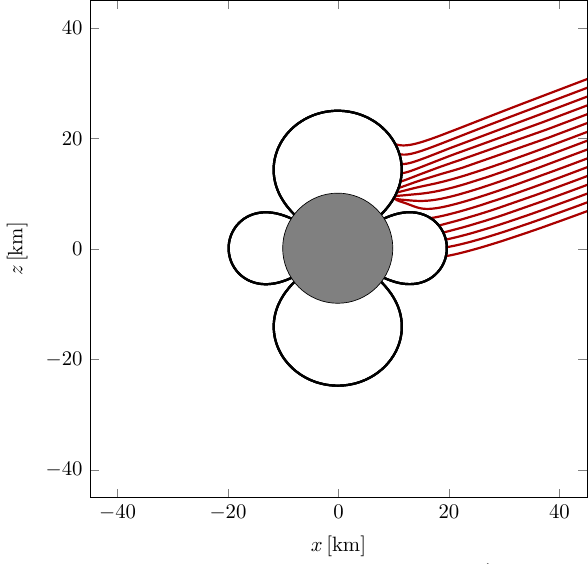}
  \includegraphics[width=0.45\textwidth]{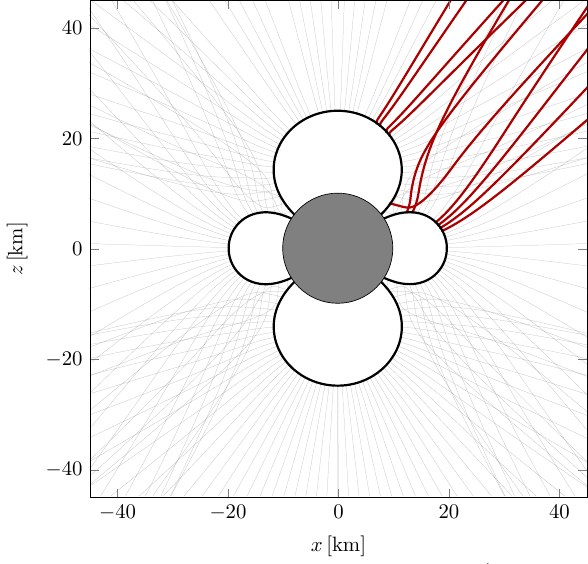}
	\caption{\textbf{Demonstration of Forward and Backward Ray Tracing.} Backward ray tracing (left) and forward ray tracing (right) procedures. We show the Langmuir-Ordinary (LO) modes for a Goldreich-Julian plasma density in a strong magnetic field plasma with dispersion relation of Eq.~\eqref{eq:LOEnergy}. Rays emanate from a plasma isosurface $\omega_{\rm p} = 10^{-5}{\rm eV}$. In the case of backward ray tracing, we show an image plane at $\theta = 1.2$ whilst in forward-tracing gray trajectories show all rays sampled from the isosurface, whilst red rays denote those binned into a viewing angle of $\theta = 0.7$. Other values were $B=10^{14}{\rm G}$, $m_a = 10^{-5}{\rm eV}$, $\alpha = 0$, $P = 2\pi$, $M_{\rm NS} = 1 M_{\rm \odot}$.
	}
	\label{fig:Rays}
\end{figure*}

 The goal of this manuscript is to develop a generalized ray tracing framework capable of treating the production of radio photons from axions in astrophysical plasma, with a particular focus on the treatment of ray tracing through a highly magnetized plasma in curved spacetime, as is required for axion searches near neutron stars. To date, the ray tracing algorithms used in this field have only incorporated a subset of the relevant effects, including either the presence of an anisotropic plasma~\cite{Witte:2021arp} or curved spacetime effects in an isotropic plasma~\cite{Battye:2021xvt}. Moreover, the methodology of these two algorithms differ markedly, with one approach back-propagating photons from an asymptotic observer, and the other forward propagating photons from the point of production (see Fig.~\ref{fig:Rays} for an illustrative example of the two setups). This work serves to unite these frameworks, highlighting a variety of important subtitles in the field which have thus far gone overlooked. While the focus of this paper will be directed toward applications of ray tracing for axion searches, the formalism is sufficiently general to be applicable, and of interest for, the broader astronomy and astrophysics communities.

The structure of this paper is organized as follows. In Sec.~\ref{sec:raytrace} we present an overview of geometric ray tracing. In particular, we derive a general formalism in which asymptotic observables (such as the radiated power in a given emission direction from a source) can be reconstructed by either $(i)$ backward propagating rays from an infinitesimal surface element located far away from the emitting region, or $(ii)$ forward propagating rays from the emitting region to asymptotic distances. This formalism can be straightforwardly applied to arbitrary spacetime metrics and any background medium -- the generalization of these techniques to photons and axions in magnetized plasmas and to curved spacetime are the focus of Secs.~\ref{sec:AxionsPlasma} and~\ref{sec:curvedS}. In Sec.~\ref{sec:NS}, we apply the formalism developed in the preceding sections to the specific problem of searching for radio spectral lines produced from the resonant conversion of axion dark matter in the magnetospheres of neutron stars. We investigate the interplay of a number of important effects, including, \eg, the impact of plasma anisotropy, strong gravity, multiply reflected photons, and a proper kinematic matching of the resonance. We also revisit the extent to which radio observations can be used to search for axions near the galactic center magnetar SGR J1745-2900; here, we apply an improved modeling of the charge distribution in the magnetosphere, showing that expected $e^{\pm}$ densities found in the closed magnetic field lines of magnetars shift the expected signal to higher frequencies, potentially generating signals in the  $\mathcal{O}(100)$ GHz - THz regime. Using rough rough estimates of the magnetic field and charge normalization, we show that an improved understanding of the properties of SGR J1745-2900 allow current and future radio and sub-mm telescopes to probe  unexplored regions of the axion parameter space. We conclude in Sec.~\ref{sec:conclusions}.

\section{Ray Tracing}\label{sec:raytrace}

In this section, we outline the basics of ray tracing required to describe radiative transport in arbitrary media. We introduce Hamilton's equations, which allow one to identify the wordlines of photons, and discuss distribution functions of photons and their phase-space integrals, which allow for the computation of observables such as energy flux or radiant intensity. We use two distinct methods to numerically calculate observational signatures relevant for astronomy. The first approach uses image-based, backward ray tracing from the observer \cite{Battye:2021xvt} and the second implements a forward ray tracing routine which directly samples the phase-space of photons produced at source, and reconstructs asymptotic observables by propagating these photons to large distances~\cite{Witte:2021arp}.

When discussing radiative transport, we are typically interested in radiation whose wavelength is much smaller than characteristic variational scales of the medium through which the radiation propagates. In that case, geometric optics applies, and one can describe photons in terms of their local phase-space coordinates given by position $x^\mu$ momentum $k_\mu$.

The dispersion relation of these photons is then described by setting a Hamiltonian $\mathcal{H} = \mathcal{H}(k , x)$ to zero $\mathcal{H} =0$, which gives a relation $k_0 = k_0(\textbf{k},\textbf{x},t)$ describing the energy of the mode. This defines a family of curves $(k_\mu(\lambda),x^\mu(\lambda))$ in phase-space,  where $\lambda$ is the wordline-parameter on such curves, along which $\mathcal{H} $ is everywhere vanishing: \ie $d \mathcal{H} (k(\lambda),x(\lambda)) /d \lambda  = 0$. By applying the chain-rule, one can see that
\begin{equation}
	\frac{d \mathcal{H} }{d \lambda} =  \frac{d k_\mu }{d \lambda}  \frac{ \partial \mathcal{H} }{\partial k_\mu}   +
	\frac{d x^\mu }{d \lambda}  \frac{ \partial \mathcal{H} }{\partial x^\mu}  =0,
\end{equation}
which is satisfied provided $x^{\mu}$ and $k_\mu$  obey
\begin{equation}\label{eq:Hamilton}
   \frac{d x^\mu }{d \lambda}  = \frac{ \partial \mathcal{H} }{\partial k_\mu}, \qquad  \frac{d k_\mu }{d \lambda}  = -  \frac{ \partial \mathcal{H} }{\partial x^\mu}\,.
\end{equation}
Eqs.~\eqref{eq:Hamilton} are the well-known Hamilton's equations, which allow one to determine the trajectory of photons from the source to the observer.

In order to connect the properties of the source (\eg emissivity) with the asymptotic observables (\eg observed flux), we can introduce Boltzmann's equation \cite{1975Hadrava,McDonald:2023ohd}
\begin{equation}\label{eq:Flasov}
\partial_k \mathcal{H} \cdot \partial_x f_\gamma  - \partial_x \mathcal{H}  \cdot \partial_k f_\gamma  = \mathcal{S}[k,x] \, ,
\end{equation}
where $f= f(k,x)$ is the phase-space distribution of photons and $\mathcal{S}[k,x]$ is a source term whose precise structure depends on the emission process.
Here, one can see that Hamilton's equations (Eq.~\eqref{eq:Hamilton}) give the characteristics (or \textit{orbits}) of the Liouville-Vlasov operator appearing on the left-hand side of Eq.~\eqref{eq:Flasov}. Thus, along worldlines, we have
\begin{eqnarray}\label{eq:BoltzWorldine}
	\frac{d f_\gamma}{d \lambda} = \mathcal{S}[\lambda]\, ;
\end{eqnarray}
in the absence of collisions ($\mathcal{S} = 0$), this implies that $f_\gamma$ is conserved along rays, \ie
\begin{eqnarray}\label{eq:sourceless}
	\frac{d f_\gamma}{d \lambda} = 0 \, ,
\end{eqnarray}
which is equivalent to Liouville's theorem \cite{1975Hadrava}. Equation \eqref{eq:sourceless} is the key equation in numerical ray tracing routines; it allows the  asymptotic distribution of photons to be reconstructed by saturating the space with rays, computing the value of $f_\gamma$ at source, and  propagating this conserved quantity along rays. In other words, ray tracing effectively amounts to solving the 8-dimensional equation \eqref{eq:Flasov}, which describes the spacetime dependence of the photon distribution function $f_\gamma$.

By integrating Eq.~\eqref{eq:Flasov}, one can also derive a continuity equation for energy \cite{McDonald:2023ohd}. Assuming a (quasi) stationary background (such that $\partial_t \mathcal{H}= 0$), Eq.~\eqref{eq:Flasov} can be pre-multiplied by $k_0$, placed on-shell, and integrated over 3-momentum $d^3 \textbf{k}$ and a spatial volume $d^3 \textbf{x} = d \mathcal{V}$. This procedure yields
\begin{equation}\label{eq:Continuity}
	\frac{d}{dt} \int d\mathcal{V} \int d^3\textbf{k} \, \omega f_\gamma + \int d^3 \textbf{k} \int d\textbf{A} \cdot \textbf{v}_g \omega f_\gamma = \int d \mathcal{V} Q \, ,
\end{equation}
where $d\textbf{A}$ is the surface element of $\mathcal{V}$ (which we assume lies outside the source), $\omega$ is the (on-shell) photon frequency, $\textbf{v}_g$ is the photon group-velocity, and $Q$ is defined by
\begin{eqnarray}\label{eq:emissivity}
 Q  \equiv   \int d^3 \textbf{k} \, \omega \, \mathcal{C}[\textbf{k},t,\textbf{x}] \, .
\end{eqnarray}
where $\mathcal{C} = \mathcal{S}/(\partial_{k_0} \mathcal{H})$ is a renormalised collision kernel whose denominator arises from dividing both sides of Eq.~\eqref{eq:Flasov} by $\partial_{k_0} \mathcal{H}$ before integrating over phase space to obtain Eq.~\eqref{eq:Continuity}. For a stationary solution, the first term on the left-hand side of Eq.~\eqref{eq:Continuity} vanishes, and the total power emanating from the source is
\begin{eqnarray}\label{eq:AsympPower}
   \mathcal{P} =  \int d^3 \textbf{k} \int d\textbf{A} \cdot \textbf{v}_g \omega f_\gamma \, .
\end{eqnarray}
Eqs.~\eqref{eq:emissivity} and \eqref{eq:AsympPower} are essential for ray tracing, as they allow for the construction of observables from the individual rays. As we will show in the following subsections, Eq.~\eqref{eq:emissivity} lies at the heart of the forward ray tracing procedure, while Eq.~\eqref{eq:AsympPower} is key to understanding backward tracing; Eq.~\eqref{eq:Continuity} serves to unite these frameworks, since for stationary solutions these two quantities are equal.

\subsection{Backward ray tracing and Imaging}

Backward ray tracing is the traditional approach used to infer the emission properties in different directions from a source. This approach proceeds by constructing an infinitesimal surface far from the object (oriented with a surface normal parallel to the line of sight), and tracing parallel rays from the surface `backwards' in time until they encounter the source itself. The differential power flowing the infinitesimal surface is given by
\begin{eqnarray}\label{eq:dp_back}
  d \mathcal{P} =   d^3\textbf{k} \,  d A  \, v_g \, \cos \theta_g \, \omega f_\gamma \, ,
\end{eqnarray}
where $\theta_g$ is the angle between the photon group velocity and the surface. Note that if the infinitesimal surface is sufficiently far from the source then one can safely assume the photon to be in vacuum, implying  $v_g = 1$, $\cos \theta_g = 1$, and  $d^3 \textbf{k} = d\Omega \, \omega^2 \, d \omega$. Following \cite{Leroy:2019ghm,Battye:2021xvt},  the power per solid angle per unit frequency is then given by a summation over each of the rays $i$, appropriately weighted by their contribution to the surface area $dA_i$, their energy $\omega_i$, and their phase space factor $f^i_\gamma$, \ie
\begin{equation}
 \frac{d \mathcal{P}_i}{d\Omega d \omega} =  \sum_i d A_i  \, \omega_i^3 f^i_\gamma \, .
\end{equation}
The only unknown quantity in Eq.~\eqref{eq:dp_back} is $f^i_\gamma$, which is determined by defining $f_\gamma$ at the source, and using the fact that $f_\gamma$ is conserved along rays.

\begin{figure*}
	\includegraphics[width=0.49\textwidth, trim={4cm 8cm 4cm 4cm}, clip]{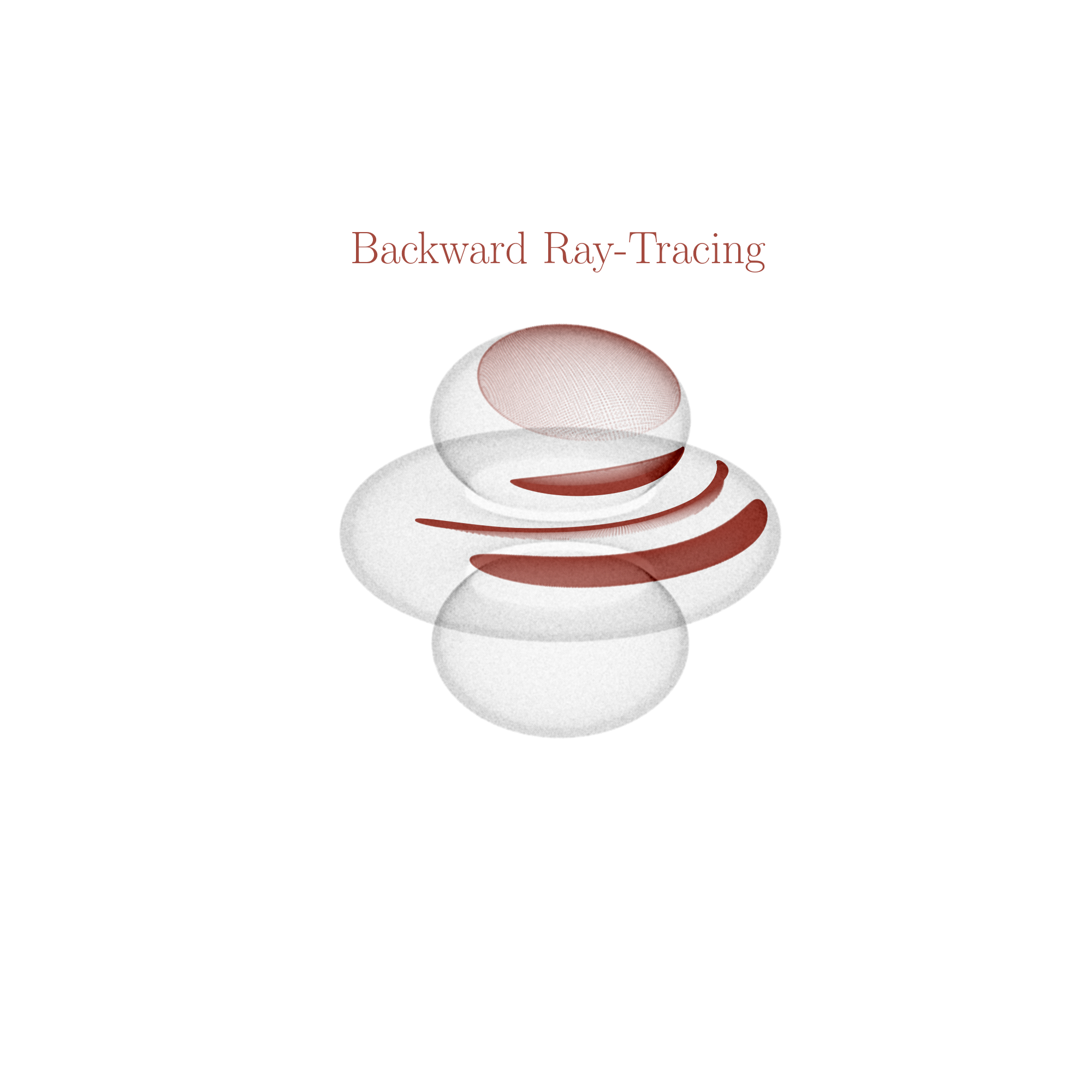}
	\includegraphics[width=0.49\textwidth,trim={4cm 8cm 4cm 4cm}, clip]{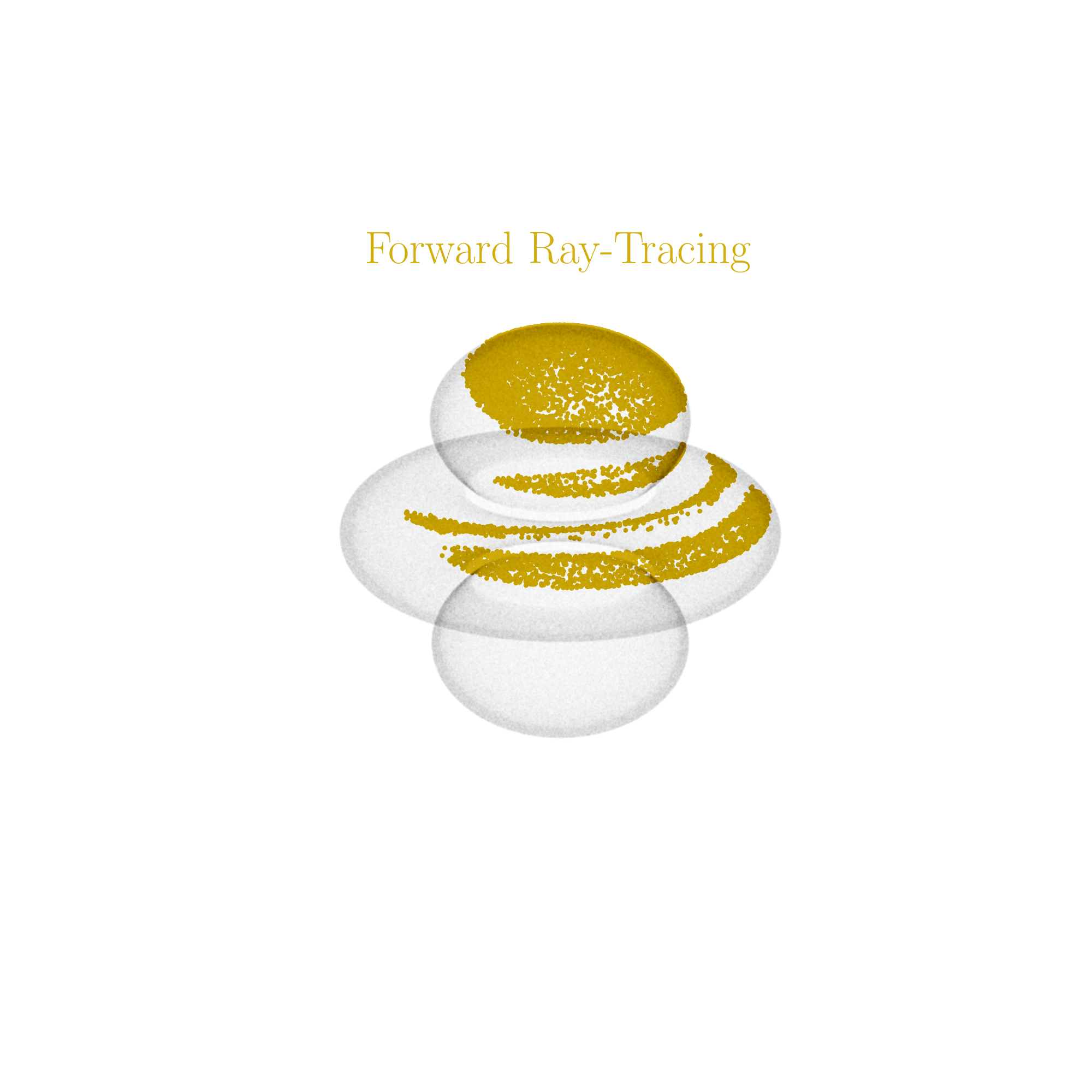}
	\caption{\label{fig:projection} \textbf{Photon Emission Points .} Subset of photon emission points from rays which propagate to an observing direction $(\theta, \phi) = (35^\circ, 0^\circ)$, projected onto the point of photon production, for the case of backward ray tracing (left) and forward ray tracing (right). Gray contours illustrate the 2D surface around the neutron star defined by $m_a = \omega_p$. Results are shown for the case of an anisotropic dispersion relation in curved spacetime, with $M_{\rm NS} = 1$ and $m_a = 10^{-5}$ eV.  }
\end{figure*}


In order to implement this procedure numerically, we define the asymptotic surface to be a plane consisting of square pixels with side length $\Delta b$. Rays are sourced through the center of each pixel, labelled by $(i,j)$, in a direction perpendicular to the surface itself. For simplicity, we assume photons are monochromatic, which allows one to evade sampling over energy. This procedure generates a differential power given by
\begin{eqnarray}\label{eq:BackPower}
	\frac{d \mathcal{P}(\theta,\varphi)}{d \Omega d\omega} = \sum_{i,j} \Delta b^2 \omega^3 f_\gamma^{i,j} \, .
\end{eqnarray}
Note, however, that non-uniform sampling of the asymptotic surface can, in general, dramatically expedite the numerical calculation.

\subsection{Forward ray tracing}
An alternative approach to backward ray tracing is to directly simulate photon production from the collision integral Eq.~\eqref{eq:emissivity}, and trace the rays \textit{forward} to all parts of the sky surrounding the source.  For a (quasi) stationary background, the power flowing at infinity is simply equal to the power produced at the source, \ie
\begin{equation}\label{eq:Power}
	\mathcal{P} = \int d^3 \textbf{x} \int d^3 \textbf{k} \,\, \omega \, \mathcal{C}[\textbf{k},t, \textbf{x}]  \, .
\end{equation}
Rather than sampling rays from an asymptotic surface (as in backward ray tracing), forward ray tracing works by stochastically sampling the photon phase space at the source, \ie it uses Monte Carlo integration to directly compute the right hand side of Eq.~\eqref{eq:Power}.
Photons are then propagated away from the source and binned on a sphere at infinity; the angular power distribution in a direction $(\theta, \phi)$ is then reconstructed by summing over the weighted rays which end up in a small bin centered about $(\theta, \phi)$.

 At this point, one may be concerned by the fact that Eq.~\eqref{eq:Power}  relates the integrated power locally to the integrated power at infinity, while the forward ray tracing procedure described above relies on the fact that this connection between local and asymptotic power also holds at the differential level.
 Notice that from Liouville's theorem, it follows that the number of particles in a phase-space element is conserved along rays, so that $dN = d^3 \textbf{k} d^3 \textbf{x} f$ is constant along rays. More explicitly, we can write this as $dN = d^3 \textbf{k} \cos \theta_g v_g dt dA$, so that the power flowing through an infinitesimal surface at the point of emission, is equal to the power flowing out of another surface at infinity (so long as these points are connected by a ray). It is ultimately for this reason that the forward propagation approach is valid, since all rays in a given part of the sky conserve the phase-space element. An equivalent way of viewing this is via the conservation of etendue. This means that not only is the total integrated power conserved (and invariant under a re-partitioning of the Monte-Carlo integration), but any two \textit{sub}-surfaces connected by rays, also have conserved power flowing between them, and are therefore also invariant under a re-partitioning of the Monte Carlo integration.

In order to provide an illustrative example of the forward ray tracing procedure, consider the case of uniform isotropic emission from a finite volume $\mathcal{V}_C$. By drawing $N_s$ uniform samples over $d^3\textbf{x}$ and $d^3\textbf{k}$, one can write the differential power flowing through a patch of the sky $(\theta_0 \pm \epsilon_\theta, \phi_0 \pm \epsilon_\phi)$ as
\begin{eqnarray}
\mathcal{P}_{\theta, \phi} \simeq   \frac{1}{N_{\rm s}} \sum_{i}   \omega_i \mathcal{C}(\vec{x}_i, \vec{k}_i) \, \mathcal{D}(\theta_{f, i}, \theta_0, \epsilon_\theta) \, \mathcal{D}(\phi_{f, i}, \phi_0, \epsilon_\phi) \, ,
\end{eqnarray}
where $\theta_{f, i}$ and $\phi_{f,i}$ are the final angular coordinates of the photon after propagation, and we have defined the function
\begin{equation}
	\mathcal{D}(x, y, \epsilon) = \begin{cases}
1 \hspace{.3cm} & {\rm if} \, y-\epsilon  \leq x \leq y + \epsilon \\
0 \hspace{.3cm} & {\rm else}
	\end{cases} \, .
\end{equation}
As in the case of backward propagation, the choice of uniform sampling may not be optimal, and one may instead prefer to implement importance sampling.

As illustrated by this example, the forward-tracing method is fully general and can be used to solve the radiative transport problem in any setup. Ultimately, however, our purpose is to study the production of photons from axions; this problem is more subtle in that axion-photon mixing is a one-to-one process, meaning that it only occurs on particular surfaces in phase-space (namely at locations where the dispersion relations of axions and photons become degenerate).  This condition collapses Eq.~\eqref{eq:Power} to a momentum-weighted sum over kinematic \textit{surfaces} rather than volumes. In the following sections we provide a more detailed discussion on how to generalize the Monte Carlo integration procedure to the case of axions.

\subsection{Comparing Forward and Backward Ray Tracing}

\begin{figure}
	\centering
	\includegraphics[width=0.45\textwidth]{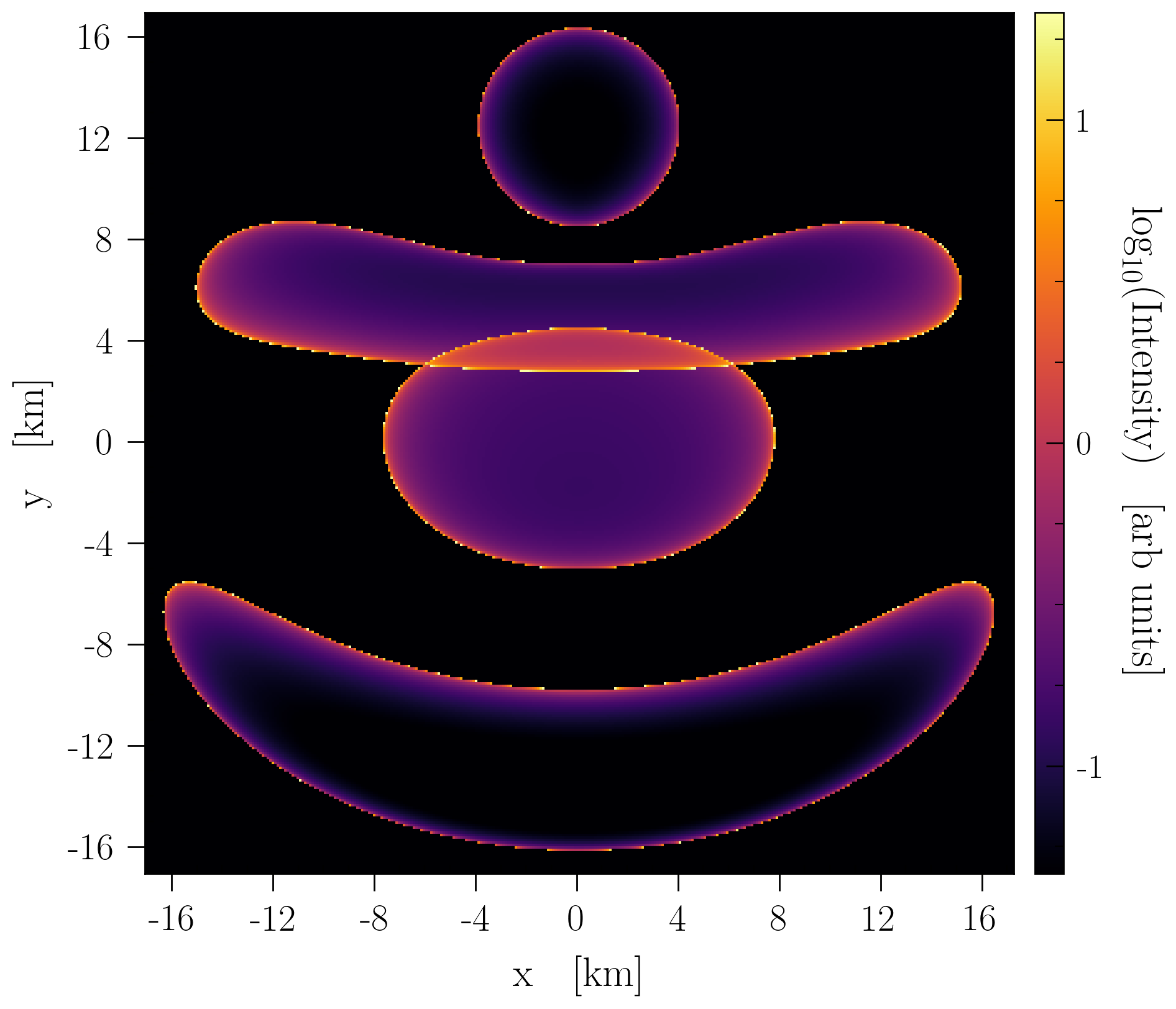}
	\caption{\textbf{Imaging in Backward Ray Tracing.} Image plane from backward ray tracing. Displayed is the intensity (energy per unit area, per unit time) flowing through the image plane in the direction $\theta = 35 ^\circ$ for $m_a = 10 \mu {\rm eV}$ in the Goldreich-Julian plasma model \cite{Goldreich:1969sb}. See Secs.~\ref{sec:AxionsPlasma}-\ref{sec:NS} for more details. }
	\label{fig:ImagePlane}
\end{figure}

Having summarized the general approach to forward and backward ray tracing, we outline here the potential advantages and disadvantages of each.
\begin{itemize}
\item Forward propagation inevitably generates samples across the entire sky (\ie the sphere at infinity surrounding the source). In the event that the observing angle is known, \ie one is only interested in the power radiated in a particular direction on the sky, the forward propagation approach clearly suffers from oversampling, since only a fraction of the samples are actually used in the computation of the relevant observable. In the event that this sampling is uniform, the over sampling may be severe, however there exist many forms of adaptive sampling algorithms which can be used to improve the sampling efficiency. This is not a problem for backward propagation, since by construction all samples originate from the angle of interest. The issue of oversampling can be seen in the comparison provided in Fig.~\ref{fig:projection}; in this example, low-energy photons are assumed to be sourced from axions near a neutron star (see Sec.~\ref{sec:AxionsPlasma}), and their point of origin for a particular viewing angle (taken here to be $(\theta, \phi) = (35^\circ, 0^\circ)$) is reconstructed using both backward and forward ray tracing techniques. While the rays originate from the same location near the neutron star, the density of samples in the forward ray tracing approach is significantly reduced with respect to the backward ray tracing example.  
\item Conversely, generating full sky distributions of the flux is more complicated in the context of backward propagation, as one must scan over (and interpolate between) many viewing angles (whereas this is a natural output of the forward sampling procedure). Full sky distributions can prove useful for understanding the fundamental physics and observables in the problem at hand, extracting the time profile of the signal, and in some cases, full sky distributions are required in order to marginalize over uncertainties associated with an unknown viewing angle. This could be circumvented by sampling the observing directions and positions on the observing plane (see Ref.~\cite{Battye:2021xvt}) stochastically (\ie, via Monte Carlo integration) rather than deterministically.

\item Backward propagation also allows one to reconstruct physical images - see Fig~\ref{fig:ImagePlane}. In this work, we are concerned principally with neutron stars, which are too small to resolve and so the precise structure of the image plays no role from a data analysis perspective in that context, since at low resolution, one is sensitive only to the integrated power over the image plane. While image reconstruction could become important in other contexts, it is unclear whether any additional information is added that cannot be directly obtained from the photon source locations (which is information that is available in both approaches).

\item As will be shown shortly, divergences arise in the conversion probability for axion-photon mixing. These divergences are naturally regulated by the phase-space measure (see discussion around Eq.~\eqref{eq:ForwardTracingEquation1}). The removal of these divergences in the forward tracing is therefore somewhat tautologous, as the cancellation occurs as soon as the integral Eq.~\eqref{eq:ForwardTracingEquation1} is written down. This integral (which is then Monte Carlo sampled in the forward-tracing numerical routine) is therefore trivially convergent. However, in backward ray tracing, the conversion probability leads to quantities which diverge at an individual ray-level; in principle, these divergences should be  regulated by the phase-space measure in the image plane. Backward ray tracing therefore provides a powerful consistency check for the treatment of phase space, kinetic theory and the conversion probability in this work and \cite{McDonald:2023ohd}, and indeed the equivalence and finiteness of forward and backward results is ensured by the expression Eq.~\eqref{eq:Continuity}. The existence of two independent numerical routines has therefore proved extremely fruitful in developing both numerical and analytic understanding of the problem at hand.

\item  Gravity plays a stabilizing role in backward ray tracing, since rays must be back-propagating from regions of high refractive index, to regions where rays are almost evanescent ($\omega_{\rm p} \simeq \omega$). This has a tendency to deflect rays away from the production surface. In the absence of gravity, ultra-high numerical precision is therefore needed for rays to converge to the point of resonance where $\omega_p  \sim m_a \simeq \omega$. In the case of forward propagation this problem is evaded, allowing one to independently study the impact of gravity on ray propagation.
\end{itemize}

\subsection{Revisions, Theory and Code Testing}

In this section, we briefly outline some of the changes which have been made both to our presentation and understanding of the theory concerning ray tracing methods since previous work \cite{Battye:2021xvt,Witte:2021arp}. We also describe revisions and improvements to the underlying codes. The backward ray tracing code is implemented in Mathematica, whilst the forward ray tracing code uses Julia, with both requiring parallelization on computing clusters.

We have carried out extensive tests of the methods used in this paper. We have verified at a ray-by-ray level that the two codes produce the same photon worldlines according to Eqs.~\eqref{eq:Hamilton} by backward propagating rays with initial conditions inferred from forward propagation from the source. We carried out this comparison for two equivalent forms of the dispersion relation presented in \cite{Turimov:2018ttf} and \cite{GedalinMelrose}. Note that two important realizations emerged in attempting to generate agreement at the ray-by-ray level. First, fundamental constants (such as the speed of flight) need to be defined equivalently to a high level of precision, and second, the image plane in the backward ray tracing approach needs to be placed at sufficiently large distances such that the vacuum approximation is valid (which is an order of magnitude larger than that used in the original work of ~\cite{Battye:2021xvt}). In addition, high-precision ODE solvers must be applied in both cases.

We checked that rays which escape to a particular direction $(\theta, \varphi)$ in the sky emanate from the same sourcing regions as those given by back-propagating from that same part of the sky (see Fig.~\ref{fig:projection}).  We confirmed that the total power computed from forward propagation (which is trivially consistent with Eq.~\eqref{eq:Power}) is equivalent to the power inferred by backward tracing to within a few percent. This was also done across a full range of observing angles, and for different photon frequencies.

We have also verified mathematically, using analytic expressions found in \cite{Rogers:2015dla,Battye:2021xvt}, that, for an isotropic plasma and a spherical emission surface, the analytic expressions Eq.~\eqref{eq:BackPower} for the power from backward tracing, and Eq.~\eqref{eq:Power} for forward tracing, are equal. Note we even verified this in curved spacetime for a Schwarzschild metric.

We also now implement correct kinematic matching of axions and photons at $k_a  = k_\gamma$, and incorporate a variety of curved-spacetime effects in each code. In addition, we make use of the new conversion probability derived in Ref.~\cite{McDonald:2023ohd}. In particular, the methods implemented here provide a much deeper understanding of phase space, and verify numerically that the divergence in the probability is regulated by phase space measure.

In the backward ray tracing code, we have also implemented an improved event-location routine to identify where photons are produced. Additionally, we now choose to describe ray tracing in terms of the photon distribution function $f_\gamma$, rather than the radiant intensity $I_\gamma$ of the photon (as was done in previous work \cite{Battye:2021xvt}). These are equivalent and related by $f_\gamma = I_\gamma /(\omega^3 n_r)$, where $n_r$ is the ray-refractive index described in \cite{Befki1966}. In particular, our discussion of phase-space in this work and Ref.~\cite{McDonald:2023ohd} now allows us to properly understand the connection between the forward and backward ray tracing methods of \cite{Battye:2021xvt} and \cite{Witte:2021arp}. As discussed in the sections below, the original work of \cite{Battye:2021xvt} did not include the possibility of photon reflection -- ray tracing was terminated when the first stopping condition, $m_a = \omega_p$, was encountered. This can prove problematic for photons sourced near the neutron star surface, since a sizable fraction of these photons undergo rapid oscillations between large plasma gradients. This has now been corrected by tracing photons for a much longer period of time, ensuring they have entered, and then escaped, the magnetosphere. Finally, in the backward ray tracing code we now implement a coupled set of first order ODEs in Eq.~\eqref{eq:Hamilton} rather than a single second order ODE given by Eq.~(20) of Ref.~\cite{Battye:2021xvt}, which is only possible in an isotropic medium.

The forward ray tracing code of~\cite{Witte:2021arp} also included one notable update of the surface sampling procedure, which now allows one to appropriately treat the resonance condition (defined by the location where $k_\mu^\gamma = k_\mu^a$, rather than $\omega_p = m_a$, with the plasma frequency defined as $\omega_{p} ^2 \equiv 4 \pi n_e /m_e$). Note that the modified resonance condition implies that photons are not sourced by a single surface, but rather a foliation of surfaces, each being locally defined by the relative orientation of the axion momentum and the magnetic field; as such, the new Monte Carlo sampling procedure picks out a single surface within the foliation by first defining the orientation of $\hat{k}_a$, and then proceeding to uniformly sample the associated two-dimensional surface (as outlined in~\cite{Witte:2021arp}).

\section{Axion-Photon Conversion and Magnetized Plasmas}\label{sec:AxionsPlasma}

We turn now to the application of ray tracing to the production of photons from axions in magnetized plasmas. In Ref.~\cite{McDonald:2023ohd}, it was shown that the axion collision integral is equivalent to
\begin{align}\label{eq:AxionCol}
C[\textbf{k}_\gamma, \textbf{x}] = & \frac{1}{\partial_{k_0} \mathcal{H}} \int \frac{d^3 \textbf{k}_a}{(2 \pi)^3 2 E_a}  (2\pi)^4\delta\left(E_\gamma(\textbf{k}_\gamma,\textbf{x}) - E_a(\textbf{k}_a) \right) \nonumber \\  & \times \delta^{(3)}\left(\textbf{k}_\gamma - \textbf{k}_a\right) \left| \mathcal{M}_{a\rightarrow \gamma} \right|^2 f_{a}(\textbf{k}_a,\textbf{x})  .
\end{align}
where $\left|\mathcal{M}_{a \rightarrow \gamma}\right|^2$ is an effective squared-matrix element for axion-photon conversion, $f_a$ is the phase-space distribution of axions, and the axion energy $E_a$ is defined by $E_a^2 = \textbf{k}^2 + m_a^2$. Note the 3-momentum integral can be performed trivially so that upon substituting Eq.~\eqref{eq:AxionCol} into Eq.~\eqref{eq:Power} we find the power is defined as
\begin{align}\label{eq:PowerExpression}
	\mathcal{P} = \int d^3\textbf{k} \int d^3 \textbf{x} \, \delta(E_\gamma(\textbf{k},\textbf{x}) - E_a(\textbf{k}))  \frac{ \pi \left| \mathcal{M}_{a\rightarrow \gamma} \right|^2 }{\partial_{k_0} \mathcal{H} }f_{a}.
\end{align}
Using the integral identity for any smooth function $G$
\begin{equation}\label{eq:IntegralTrick}
	\int_{\mathbb{R}^n} d^n \textbf{z} \, \delta(G(\textbf{z})) = \int_{G^{-1}(0)} d\Sigma  \,\frac{1}{\left| \nabla G \right|},
\end{equation}
where $d\Sigma$ is the area element on the surface defined by $G(\textbf{z}) =0$, the integration of the delta function in Eq.~\eqref{eq:PowerExpression} can then be performed with respect to the spatial integral, giving

\begin{align}\label{eq:AxionCollInt}
\mathcal{P}  =
\int  d^3 \textbf{k} \int  d \Sigma_\textbf{k}  \frac{\pi \left| \mathcal{M}_{a\rightarrow \gamma} \right|^2}{  \partial_{k_0} \mathcal{H}\left| \nabla_\textbf{x} E_\gamma\right| } f_{a} \, ,
\end{align}
where $\Sigma_\textbf{k}$ is a surface defined by the set of points
\begin{equation}\label{eq:Hypersurfaces}
 \Sigma_\textbf{k} = \left\{ \textbf{x} : E_a(\textbf{k},\textbf{x}) - E_a (\textbf{k} ) = 0 \right\} \, ,
\end{equation}
so that over a continuum of $\textbf{k}$ values, the $\Sigma_\textbf{k}$ define a foliation of distinct surfaces.

As a special case, note that for an isotropic (unmagnetized) cold plasma, we have and $E_\gamma = \textbf{k}^2 + \omega_p^2$ so that the surface on which the energies become degenerate is given simply by $\omega_p = m_a$, in which case the foliation collapses to a single surface $\Sigma$, independent of $\textbf{k}$. In general, however,  the dispersion relation (and hence $E_\gamma$) can also depend on the direction of $\textbf{k}$, generating a distinct surface for each value of the momentum; as described below, this is the case in a strongly magnetized plasma, as found near the surfaces of neutron stars.

Returning to the expression Eq.~\eqref{eq:AxionCollInt} for the axion collision integral, we can then insert a factor of $v_a$ and $\cos \theta_{\textbf{n}}$ in the denominator and the numerator, where $\cos \theta_{\textbf{n}}$ is the cosine of the angle between the phase velocity $\textbf{v}_p = \textbf{k}/E_\gamma$ and the normal to the surface surface $\Sigma_\textbf{k}$. Here $\textbf{v}_p$ is the phase velocity of the axion, equal to that of the photon at the resonant surface. We then also use the relation $\Sigma_\textbf{k} \cos \theta_\textbf{n} v_p = d\boldsymbol{\Sigma}_p \cdot \textbf{v}_p$ and $v_g \cos \theta_\textbf{n} \left| \nabla_\textbf{x} E_\gamma \right| = \left|\textbf{v}_p \cdot \nabla_\textbf{x} E_\gamma \right| $. This allows us to write
\begin{eqnarray}\label{eq:ForwardTracingEquation1}
	\mathcal{P} = \int d^3 \textbf{k} \int d  \boldsymbol{\Sigma}_\textbf{k}\cdot \textbf{v}_a P_{a \gamma} \, \omega f_a.
\end{eqnarray}
where
\begin{eqnarray}
	P_{a \gamma } = \frac{\pi \left| \mathcal{M}_{a\rightarrow \gamma} \right|^2}{ E_\gamma  \partial_{k_0} \mathcal{H} \left| \textbf{v}_a\cdot \nabla_\textbf{x} E_\gamma  \right|}
\end{eqnarray}
is a conversion probability, which, as shown in Ref.~\cite{McDonald:2023ohd}, defines the ratio of axion to photon phase-space densities at the resonance:
\begin{eqnarray}\label{eq:ProbFlat}
	P_{a \gamma \gamma}(\textbf{k},\textbf{x}) \equiv \frac{f_\gamma(\textbf{k},\textbf{x})}{f_a(\textbf{k},\textbf{x})}.
\end{eqnarray}

The general form of $P_{a \gamma}$ of course depends on the medium in question. For strongly magnetized plasmas (as is relevant for this work), the photon mode of interest is the Langmuir-Ordinary (LO) mode (see~\cite{Witte:2021arp}); in non-relativistic plasmas, the energy of the LO mode satisfies
\begin{equation}\label{eq:LOEnergy}
	E_\gamma^2 = \frac{1}{2} \Bigg[ \omega_p^2 + k^2 + \sqrt{k^4 + \omega_p^4  + 2 k^2 \omega_p^2 (1 - 2 \cos^2 \theta_B}) \Bigg],\,
\end{equation}
where $\theta_B$ is the angle between $\textbf{B}$ and $\textbf{k}$.
Using this expression for the photon energy, one finds a conversion probability given by \cite{McDonald:2023ohd}
\begin{equation}\label{eq:CPStrongB}
    P_{a \gamma } = \frac{\pi}{2}\cdot \frac{ g_{a \gamma \gamma}^2\left| \textbf{B}_{\rm ext} \right|^2E_\gamma ^4 \sin ^2 \theta }{\cos ^2 \theta
   \omega _p^2 \left(\omega _p^2-2 E_\gamma ^2\right)+E_\gamma ^4} \cdot    \frac{1}{\left| \textbf{v}_a \cdot \nabla_\textbf{x} E_\gamma  \right|} .
\end{equation}

Eq.~\eqref{eq:ForwardTracingEquation1} thus relates the power $\mathcal{P}$ flowing through a bounding surface $A$ to a term involving an integration over axion phase-space at the source. The surface $A$ can then be chosen to be the sphere at infinity, thus relating the power at source to the power measured by observers far away.

As a reminder, the general idea behind the forward propagation approach is to directly compute the right hand side of Eq.~\eqref{eq:ForwardTracingEquation1} locally at the source using Monte Carlo integration, and then use ray tracing to relate the local properties of emission to the asymptotic properties by exploiting the fact that $f_\gamma$ is conserved along trajectories once outside the source (Eq.~\eqref{eq:sourceless}). Alternatively, the backward ray tracing approach tracks the conserved phase-space density $f_\gamma$ along photon world-lines until $k_\gamma= k_a$, at which point one assigns a value $f_\gamma = P_{a \gamma} \times f_a$. Note that in the case of backward ray tracing, trajectories must be traced until they have passed through and fully escaped the gravitational potential of the neutron stars -- this is a consequence of the fact that each photon worldline can encounter multiple level crossings, and the total photon phase space is the weighted sum of each of these.


\section{Generalization to   Curved Spacetime} \label{sec:curvedS}

In this section, we discuss how to generalize the results presented in previous sections to curved spacetime. We focus in particular on the case of photon production via axions in strongly magnetized plasmas in the presence of gravitational fields. This is particularly relevant in the vicinity of neutron stars, which are the most compact stars in existence, with the ratio of the Schwarzschild radius to the neutron star radius being $r_s/R \sim 0.3$.

The importance of incorporating gravitational effects is multi-faceted. Firstly, gravity alters photon trajectories relative to flat spacetime, acting as an attractive force which counteracts repulsive refraction due to plasma gradients~\cite{1975Hadrava,GedalinMelrose,Rogers:2015dla,Battye:2021xvt}. Next, axions fall towards the star along geodesics of the stellar spacetime metric. Energy-momentum conservation demands that axions and photons should have matching 4-momenta at the point of creation, $k^a_\mu = k^\gamma_\mu$. For kinematic self-consistency, one should therefore demand that gravitational effects are incorporated self-consistently in the axion and photon dispersion relations. By doing so, one changes the geometry of the foliation $\Sigma_\textbf{k}$ in Eq.~\eqref{eq:ForwardTracingEquation1} relative to flat space, with the innermost and outermost surfaces in the foliation separated by a greater distance than in flat space.  In addition, metric contributions will enter the infinitesimal area $d\Sigma_\textbf{k}$ appearing in the collision integral.

A related consequence is the fact that gravity increases the density of axion near the star \cite{Hook:2018iia} -- this arises because geodesic bundles of axions become focused in regions of strong gravity (causing a greater number of axions to cross the resonant conversion region per unit time), and can lead to a sizable enhancement of the amplitude of the radio flux. This effect has been included in previous work via a renormalization of $f_a$, but in generalization must be done self-consistently with the aforementioned effects.

Turning to the production process itself, the analytic expression for the conversion probability in flat space is intimately connected to phase-space, kinematics, dispersion relations, and worldlines of photons and axions~\cite{McDonald:2023ohd}. Crucially, divergences emerging in the conversion probability are shown in flat space to be canceled by compensating terms in the phase-space integration in Eq.~\eqref{eq:ForwardTracingEquation1}. As a result, a fully self-consistent generalization of kinetic theory and ray tracing to curved spacetime is also relevant for properly describing the efficiency with which photons are produced.

Finally, as pointed out in \cite{Foster:2022fxn}, strong magnetic fields and larger values of $g_{a \gamma \gamma}$ can lead to $\mathcal{O}(1)$ axion-photon transitions. Computing the radio signals in this scenario requires self-consistently tracking axion and photon trajectories through all resonant crossings (as well as the trajectories of axions and photons sourced from those resonances); although we do not go beyond perturbative production in this work, the techniques developed here lay the groundwork for such follow-up analyses.

In the remainder of this section, we outline how to incorporate curved spacetime effects into each step of the problem.

\subsection{Magnetized Plasmas in Curved Spacetime}\label{subsec:MagPlasma}

Let us begin by defining the generalized Hamiltonian that describes the covariant treatment of waves in a magnetized plasma.
The covariant treatment of waves in a magnetized plasma on curved spacetime can be found in \cite{GedalinMelrose,Turimov:2018ttf, BreuerEhlers1980,BreuerEhlers1981II,BreuerEhlers1981a} (the less general case of an isotropic plasma is studied in \cite{1975Hadrava,Rogers:2015dla}). For a non-relativistic plasma, with fluid velocity $u^\mu$, phase space coordinates $(x^\mu, k_\mu)$ and background electromagnetic field strength tensor $F^{\mu \nu}$ and spacetime metric $g_{\mu \nu}$,  we find the Hamiltonian is given by
\begin{align}
&\mathcal{H}(x ,k) \nonumber \\
&= (\omega^2-p^2)\left[\omega^2\omega^2_{\rm L}(\omega^2-\omega^2_p-p^2)
+\omega^2_p(\omega_{\rm L}\cdot p)^2\right]
\nonumber \\
& -\omega^2(\omega^2-\omega^2_p)(\omega^2-\omega^2_p-p^2)^2\ \label{DisRel}.
\end{align}
where
\begin{equation}\label{eq:Projection}
	\omega  = u^{\mu} k_\mu, \qquad p = h \cdot k \, ,
\end{equation}
and
\begin{equation}
	h_{\mu \nu} = g_{\mu \nu}  + u_\mu u_\nu, \qquad u^2 = -1 \, ,
\end{equation}
with
\begin{equation}\label{eq:Larmor}
\omega_{\rm L} = \sqrt{\omega_{\rm L \, \mu}\omega_{\rm L}^{\mu}}\ ,
\quad \omega_{\rm L}^{\mu} = \frac{e}{2m_e}\varepsilon^{\mu\nu\alpha\beta}u_\nu
B_{\alpha\beta} \ ,
\end{equation}
where $B_{\alpha \beta} = h_{\alpha \mu }h_{\beta \nu} F^{\mu \nu}$, $\varepsilon^{\mu \nu \alpha \beta}$ is the Levi-Civita tensor, $h_{\mu \nu}$ is the projection operator onto the rest-frame of the plasma, $\omega_{\rm L}$ is the Larmour frequency, and $\omega$ and $p_\mu$ are the frequency and 4-momentum in the rest-frame of the plasma.

We can also define the magnetic field via \cite{GedalinMelrose}
\begin{eqnarray}
	B^\mu = \varepsilon^{\mu \nu \rho \sigma} F_{\nu \rho} u_\sigma .
\end{eqnarray}
Using these relations, we have
\begin{align}
	\omega_L^\mu &= \frac{e}{2 m} B^\mu, \\
	\omega_L \cdot p &= \omega_L \cdot k \\
	p^2 & = k^2 + \omega^2,
\end{align}
where the first two identities follow from the anti-symmetric properties of the Levi-Civita tensor when contracted with multiple $u_\mu$, which also implies $u \cdot \omega_L = 0$.  It therefore follows that the Hamiltonian can be re-written as
\begin{align}
\mathcal{H}(x ,k) =& k^2 \left[\omega^2\omega^2_{\rm L}(k^2 + \omega_p^2)
-\omega^2_p(\omega_{\rm L}\cdot k)^2\right]
\nonumber \\
&-\omega^2(\omega^2-\omega^2_p)(k^2 + \omega_p^2)^2 \, .
\end{align}
We can now go to the strong magnetization limit relevant for neutron star magnetospheres. This is characterized by the limit $\omega_{\rm L} \gg \omega, \omega_{\rm p} $. In that limit, the dispersion relation $\mathcal{H} =0$ can be equivalently realized by the Hamiltonian
\begin{equation}\label{eq:StrongBHam}
\mathcal{H} = g^{\mu \nu} k_\mu k_\nu + (\omega ^2 - k^2_\parallel)\frac{\omega_p^2}{\omega^2}
\end{equation}
where $k_\parallel = k \cdot B/\sqrt{B.B}$, which is identical to the dispersion relation in \cite{GedalinMelrose} in the limit of a non-relativistic plasma\footnote{Note some sign differences due to a different metric signature used between Refs.~\cite{Turimov:2018ttf, BreuerEhlers1980,BreuerEhlers1981II,BreuerEhlers1981a} and \cite{GedalinMelrose}. } .

We can also define an angle between $B^\mu$ and $k_\mu$
\begin{eqnarray}
	\cos \theta_B = \frac{B \cdot  k }{ k B }.
\end{eqnarray}
where $k = \sqrt{k \cdot k}$. This allows us to write $k_\parallel^2 =k^2 \cos^2 \theta_B $.

For stationary plasmas and spacetimes, we can choose the fluid to be instantaneously at rest with respect to the space-time metric. For example, for a Schwarzschild metric
\begin{eqnarray}\label{eq:SchwarzshildMetric1}
	ds^2 = - A \, dt^2 +  A^{-1} dr^2 + r^2 d\Omega^2
\end{eqnarray}
where $A \equiv (1 -  r_s / r)$ and $r_s \equiv 2 GM$, we can define generalized phase-space coordinates $(t,r,\theta,\varphi)$ and $(k_t, k_r, k_\theta,k_\varphi)$ and a fluid velocity
\begin{equation}
	u^\mu = \left(\sqrt{ - g_{tt}^{-1}}, \, 0, \, 0 \,,\, 0 \right).
\end{equation}
In this case, $\omega^2 = - g^{tt} k_t^2  $ and hence, $k^2= g^{tt} k_t^2 + g^{ij}k_i k_j = -\omega^2 + \left| \textbf{k}\right|^2 $, with $|\textbf{k}|^2 = g^{i j}k_i k_j $ and where $i,j \in (r,\theta,\varphi)$ label the spatial components of tensors. We also have $B^0 =0$. Hence in this frame, we can write the dispersion relation as
\begin{equation}\label{eq:AnisotrpoicDisp}
 \omega^2 \left( \omega^2 - |\textbf{k}|^2 - \omega_p^2 \right) + |\textbf{k}|^2 \omega_p^2\cos^2 \theta_B =0,
\end{equation}
which gives
\begin{align}\label{eq:AnisotropicDisp}
   \omega^2 =& \frac{1}{2} \Bigg[ |\textbf{k}|^2 + \omega_p^2 \nonumber \\
	&\pm \sqrt{|\textbf{k}|^4  + \omega_p^4 +2 |\textbf{k}|^2 \omega_p^2( 1 -2  \cos^2 \theta_B )} \, \, \Bigg],
\end{align}
with the `+' and `-' signs corresponding respectively to the LO mode and Alf{\'e}n mode  of Eq.~\eqref{eq:LOEnergy}. Hence, to construct rays in curved spacetime, we solve Hamilton's equation Eq.~\eqref{eq:Hamilton} with Eq.~\eqref{eq:StrongBHam}. In the next subsection, we describe how to generalize forward ray tracing to curved spacetime, focusing in particular on how to generalize the area measure $\Sigma_\textbf{k}$ appearing in Eq.~\eqref{eq:ForwardTracingEquation1}.

\subsection{Radiative Transport in Curved Spacetime}\label{Sec:curvedSpacetimem}

In this section, we discuss how to generalize the kinetic theory for the production of photons to curved spacetime. In particular, generalizing forward ray tracing to curved spacetime necessitates generalizing the expression Eq.~\eqref{eq:Power}, or, more specifically to axions, the surface integrals in Eq.~\eqref{eq:ForwardTracingEquation1}. In flat space the infinitesimal power flowing per unit time $dt$ can be written as
\begin{equation}\label{eq:dPFlat}
	d \mathcal{E} =  dt  \, d^3 \textbf{k}  \, d  \Sigma_\textbf{k} \cos \theta_\textbf{n}  v_a \omega P_{a \gamma} \, f_a,
\end{equation}
where we have explicitly written $d  \boldsymbol{\Sigma}_\textbf{k}\cdot \textbf{v}_p = d  \Sigma_\textbf{k} \cos \theta_\textbf{n} v_a$ with $v_a = \left| \textbf{k} \right|/\omega$.

To generalize this to curved spacetime, we recognize that $dt d\Sigma_\textbf{k}$ is to be interpreted as the spacetime volume element on a $2+1$ dimensional submanifold. For a metric of the form Eq.~\eqref{eq:SchwarzshildMetric1},
the temporal piece of the volume element is generalized by taking
\begin{eqnarray}
dt \rightarrow \sqrt{\left| g_{tt} \right|} dt.
 \end{eqnarray}
The generalization of $d\Sigma_\textbf{k}$ to curved spacetime is given by taking
\begin{eqnarray}
	d \Sigma_\textbf{k} = \sqrt{\left| h_\textbf{k}\right|}
\end{eqnarray}
where $\left| h_\textbf{k}\right|$ is the determinant of the pull-back metric  $h_\textbf{k}$ corresponding to embedding the surfaces in Eq.~\eqref{eq:Hypersurfaces} in the background spacetime of $g_{\mu \nu}$. The metric $h$ corresponds to a 2D spatial surface, which we can label by their coordinates $(\theta,\varphi)$ with the radial coordinate of the surface corresponding given by
\begin{equation}\label{eq:RaidalHypersurface}
	r = r_\textbf{k}(\theta,\varphi),
\end{equation}
where the $\textbf{k}$ subscript reminds us that these surfaces form a foliation parametrized by $\textbf{k}$. The $h_\textbf{k} = h_\textbf{k}(\theta,\varphi)$ can be read off from the spatial line-element $d\bar{s}^2$ of the Schwarzschild metric, defined by
\begin{eqnarray}\label{sbar}
	d\bar{s}^2 \equiv A^{-1} dr^2 + r^2 \left( d\theta^2 + \sin^2 \theta d\varphi^2\right).
\end{eqnarray}
Using the chain-rule we can project this line element onto the surfaces corresponding to Eq.~\eqref{eq:RaidalHypersurface}, giving
\begin{equation}
dr = (\partial_\theta r_\textbf{k}) d\theta + (\partial_\varphi r_\textbf{k}) d\varphi \, .
\end{equation}
Inserting this result into Eq.~\eqref{sbar}, one can directly read of the elements of $h_{\alpha \beta}$; the determinant is then given by
\begin{align}
\sqrt{|h_\textbf{k}|} &=  A^{1/2} \, r \, \sqrt{r^2 \sin^2\theta + A^{-1} ((\partial_\theta r)^2 \sin^2\theta + (\partial_\phi r)^2 ) } \, , \\
\label{eq:redshift}
	\sqrt{|g_{tt}|} &= \left(1 - \frac{r_s}{r}\right)^{1/2},
\end{align}
where, for compactness we have dropped the $\textbf{k}$ subscript on $r$.  This provides the generalization of $d \Sigma_\textbf{k}$ to curved spacetime. Notice that the expression in Eq.~\eqref{eq:redshift} gives precisely the gravitational redshift factor that accounts for energy loss of photons propagating out of the gravitational potential.

Putting everything together, we find the  power integral in curved spacetime can be expressed as
\begin{eqnarray}\label{eq:PowerCurved}
	\mathcal{P} = \int d^3 \textbf{k} \int d\theta d\varphi \sqrt{|h_\textbf{k}|} \cos \theta_\textbf{n} v_a \omega P_{a \gamma} \, f_a.
\end{eqnarray}
Notice that here $d^3 \textbf{k}$ corresponds to the volume element of any orthonormal tetrad on the manifold used to define momentum space locally. It can be rotated into any other orthonormal tetrad basis using a local SO(3) rotation, which does not change the momentum volume measure, because the determinant of the SO(3) matrix is 1. The angle $\theta_\textbf{n}$ should be defined covariantly via
\begin{eqnarray}\label{eq:AngleDef}
	\cos \theta_\textbf{n} = \frac{n^\mu k_\mu}{\sqrt{k\cdot k}}
\end{eqnarray}
where $n^\mu$ is the unit normal to the surface $\Sigma_\textbf{k}$.

\subsection{Gravitational Focusing of Axions}

In general, the asymptotic axion distribution is expected to follow a Maxwell-Boltzmann distribution with a non-relativistic momentum dispersion $k_0 = m_a v_0$ with $v_0 \sim 220$ km/s, so that
\begin{eqnarray}\label{eq:GaussianDist}
   \lim_{|\textbf{x}|\rightarrow \infty} f_{a}(\textbf{x},\textbf{k}) = \frac{n_{a,\infty}}{(\pi k_0^2)^{3/2}} \, e^{-\left| \textbf{k} \right|^2/ k_0^2} \, ,
\end{eqnarray}
where we have assumed for simplicity that the pulsar is at rest with respect to the rest frame of the asymptotic axion distribution.

To compute the phase-space density near the star, we first make use of conservation of energy,
\begin{eqnarray}\label{eq:conservation}
   \frac{ \left| \textbf{k}(s) \right|^2}{2 m_a} - \frac{GM m_a}{\left|\textbf{x}(s) \right|}= \frac{\left| \textbf{k}_\infty\right|^2}{2m_a},
\end{eqnarray}
where $\left|\textbf{k}_{\infty}\right|$ is the asymptotic momentum of the axion at infinity and $s$ is an axion worldline parameter. Note that knowledge of the asymptotic solutions, Liouville's theorem, and kinematic constraints like Eq.~\eqref{eq:conservation}, are in general \textit{not} sufficient to solve for the functional form of the distribution function globally -- that is precisely why ray tracing is employed. However, in the present context, owing to the uniform and isotropic boundary conditions, we can construct solutions using simple conservation arguments, which effectively allow us to circumvent ray tracing for axions, allowing us to determine their distribution analytically. We know, by Liouville's theorem that
\begin{align}
	f_a(\textbf{x}(s),\textbf{k}(s) )
	&=  f(\textbf{x}_\infty, \textbf{k}_\infty ) \nonumber \\
	&=  \frac{n_{a,\infty}}{(\pi k_0^2)^{3/2}} \, e^{-\left| \textbf{k}_{\infty} \right|^2 / k_0^2} \,
\end{align}
and using the energy conservation equation (\ie Eq.~\eqref{eq:conservation}) on the right-hand side, and Liovuille's theorem on the left hand-side, one has
\begin{align}
 	&f( \textbf{x}(s), \textbf{k}(s) )  \nonumber \\
 	&= \frac{n_{a,\infty}}{(\pi k_0^2)^{3/2}} \, \exp\left( -\frac{1}{k_0^2}\left[ \left| \textbf{k}(s) \right|^2 - \frac{2 GM m_a^2}{\left|\textbf{x}(s) \right|} \right] \right) .
\end{align}
Since this holds for \textit{any} ray worldlines $\textbf{k}(s)$ and $\textbf{x}(s)$, this must therefore give the \textit{global} solution
\begin{align}
&f_a(\textbf{x},\textbf{k})\nonumber \\
&= \frac{n_{a,\infty}}{(\pi k_0^2)^{3/2}} \, \exp\left( -\frac{1}{k_0^2}\left[  \left| \textbf{k} \right|^2 - \frac{2GM m_a^2}{\left|\textbf{x} \right|} \right] \right) .
\end{align}
The reason we are able to determine the distribution of axions given boundary conditions on a surface at infinity, \textit{without} needing to ray-trace, is a special case arising from the high degree of symmetry in the problem. Indeed, the solution we have constructed is nothing more than $f_a \propto \exp(-\mathcal{H}_a/k_0^2)$ where $\mathcal{H}_a (\textbf{x},\textbf{k}) = \textbf{k}^2/(2m) - 2GM m_a/\left|\textbf{x}\right| $, with the normalisation fixed by boundary conditions. This is a trivial solution to the collisionless, stationary, axion Boltzmann equation
\begin{eqnarray}\label{eq:BoltzannAxion}
	\left\{  \mathcal{H}_a ,f_a  \right\} \equiv \partial_\textbf{k} \mathcal{H}_a \cdot  \nabla_\textbf{x} f_a - \partial_\textbf{x} \mathcal{H}_a \cdot  \nabla_\textbf{x} f_a = 0 ,
\end{eqnarray}
where the left-hand side gives the Poisson bracket. Clearly the solution we have constructed satisfies this equation since $f_a = f_a(\mathcal{H}_a)$ gives a trivially vanishing poisson bracket. For more complicated distributions, such as, \eg~, the scenario in which axions are predominantly confined to gravitationally bound miniclusters, these simple solutions do not apply, and one must also apply ray tracing to the in-fall of axions, see,  \eg~,\cite{Witte:2022cjj}.

Returning to our main discussion, we see the distribution in Eq.~\eqref{eq:GaussianDist} gives axions with characteristic asymptotic momentum $\mathcal{O}(k_0)$, for simplicity, we therefore choose to evaluate the ray tracing algorithms by isolating a single momentum $k = k_0$, for axions, \ie ,
\begin{eqnarray}\label{eq:fa_approx}
  \lim_{|\textbf{x}|\rightarrow \infty} f_{a}(\textbf{x},\textbf{k}) \simeq \frac{n_{a,\infty}}{4\pi k_0^2} \, \delta\left(\left|\textbf{k}\right| - k_0\right) \, ,
\end{eqnarray}
that is, a monochromatic, zero-width distribution with characteristic momentum $k = k_0$. This procedure can be shown to yield nearly an identical density enhancement as the Maxwellian distribution\footnote{This can be computed by replacing $v_\infty \rightarrow v_\infty[v[\vec{x}]]$ (where $v[\vec{x}]$ is the local velocity at a point $\vec{x}$ along a fixed trajectory) in Eq.~\ref{eq:fa_approx},  assuming $f_a$ is conserved along rays, and integrating over $d^3 v$ at a fixed radii near the neutron star.}, but significantly reduces the sampling required in the backward ray tracing approach since only a single photon frequency must be sampled in the image plane\footnote{Note that in a non-stationary background the photon frequency could evolve notably away from the central value, and a more sophisticated frequency sampling may be required. This effect is not expected to be important in the case axions, however~\cite{Witte:2021arp,Battye:2021xvt}. }. We return in the later sections to illustrate the impact of this simplifying assumption.
Following similar arguments to above, we determine the global distribution to be
\begin{align}
  f_{a}(\textbf{x},\textbf{k}) &= \frac{n_{a,\infty}}{4\pi k_0^2} \, \delta\left(
  \left[ \left|\textbf{k}\right|^2 - \frac{2GM m_a^2}{\left| \textbf{x}\right|} \right]^{1/2} - k_0
  \right)  \, ,
\end{align}
In particular, integrating this equation over momentum space, gives
\begin{align}
  	n(\textbf{x})  &= \int d^3 \textbf{k} \, f_a(\textbf{x},\textbf{k})  \nonumber \\
  & =  n_{a,\infty} \frac{k_c(\left|\textbf{x}\right|)}{k_0}
\end{align}
where $k_c^2 = k_0^2 + 2 G M m_a^2/|\textbf{x}|$
so that the axion \textit{number} density is enhanced by the ratio of the asymptotic and escape velocities, due to focusing of the gravitational field. Putting this together, we can express the axion distribution function in terms of the local number density of axions:
\begin{eqnarray}
  	f_{a}(\textbf{x},\textbf{k}) =
  v_a n_{a,c} \frac{\delta\left( \omega - \omega_c
  \right)}{4 \pi k^2 }  \, ,
\end{eqnarray}
where $\omega_c = \sqrt{m_a^2 + k_c^2}$ and $n_c$ and $v_a$ are the axion number density and phase velocity, respectively, both evaluated at source.

One might wonder why it is that this gravitational focusing of axions is not reversed when photons climb out of the gravitational potential. The reason for this, is that photons do not experience the same refractive index as axions as they exit the potential - they are in plasma, not vacuum, and asymptote to a refractive index $n_\gamma \rightarrow 1$ as they move away from the star, whilst axions approach $n_a \rightarrow v_0$. This is especially clear in the isotropic plasma, where the distribution function $f = I/(\omega^3 n^2_r)$ is conserved along rays \cite{1975Hadrava,Battye:2021xvt} -- here, we have defined $n_r$, the refractive index, and $I$ is the radiant intensity. Using conservation of $f$ thus implies
\begin{align}\label{eq:IntensityChain}
	\frac{I_a^\infty}{\omega_{\infty}^3 (n^\infty_{r, a})^2} = 	\frac{I_a^c}{\omega_{\rm c}^3 (n_{r,a}^c)^2} &=  \frac{1}{P_{a \gamma}}\frac{I_\gamma^c}{\omega_{\rm c}^3 (n^c_{r,\gamma})^2}  \nonumber \\
	&=  \frac{1}{P_{a \gamma}}\frac{I_\gamma^\infty}{\omega_{\infty}^3 (n_{r,\gamma}^{\infty})^2} ,
\end{align}
where in the second to third equations, we used $f_\gamma = P_{a \gamma} f_a$. Putting this together, we see firstly that
$I^c_a \propto (n_{r, a}^c)^2 /(n^\infty_{r, a})^2 I_a^\infty \sim (2GM/v_0)^2 I_a^\infty \gg I_a^\infty$ so that gravitational focusing increases the intensity of axions. Secondly, we have
\begin{eqnarray}\label{eq:PhotonFocuss}
	I_\gamma^\infty = P_{a \gamma} \frac{n^2_{\gamma,\infty}}{n^2_{a,\infty}} I_a^\infty \simeq  P_{a \gamma} \frac{I_a^\infty}{v_0^2} ,
\end{eqnarray}
where we used, $n_{r, \gamma}^\infty =1$ and $n_{r, a}^\infty = v_0$. We see that (in this illustrative isotropic case) the extent to which the gravitational focusing of axions is undone as photons exit the plasma potential, is precisely captured by the ratio of the asymptotic refractive indices of the two species. We notice that by equating each of the equations Eq.~\eqref{eq:IntensityChain} and  Eq.~\eqref{eq:PhotonFocuss}, the volume  element change \cite{Battye:2021xvt} from gravitational infall of axions, $\propto (\omega_c/\omega_\infty)^3 \simeq (1 - r_s/r)^{-3/2}$, is undone when photons exit the plasma, since this term is purely gravitational. However, the ratio of the refractive indices in Eq.~\eqref{eq:PhotonFocuss} quantifies the difference of the two potentials through which axions enter and photons exit the plasma. Similar logic holds  for the anisotropic medium.

\subsection{Conversion Probability in Curved Spacetime}

The conversion probability for axion-photon conversion in 3D magnetized plasmas has recently been computed in flat space \cite{McDonald:2023ohd}, yielding the result given in Eq.~\eqref{eq:CPStrongB} (an expression which is valid
in a (quasi) stationary background).  Formally, given we are including curved spacetime effects in our analysis, self-consistency implies we should also generalize the production rate in Eq.~\eqref{eq:CPStrongB} to curved spacetime.

One important reason for doing so is that divergences in the conversion probability should be regulated by the phase-space measure. More specifically, in flat space the surface normal of $\Sigma_\textbf{k}$ is parallel to $\nabla_\textbf{x} E_\gamma$, so that divergences occurring when the phase-velocity $\textbf{v}_a$ is perpendicular to $\nabla_\textbf{x} E_\gamma$ are regulated by the flux-like projection $\textbf{v}_p \cdot \boldsymbol{\Sigma}_\textbf{k}$, where $\boldsymbol{\Sigma}_\textbf{k}$ is the directed surface element.

\begin{figure}
\includegraphics[width=0.49\textwidth]{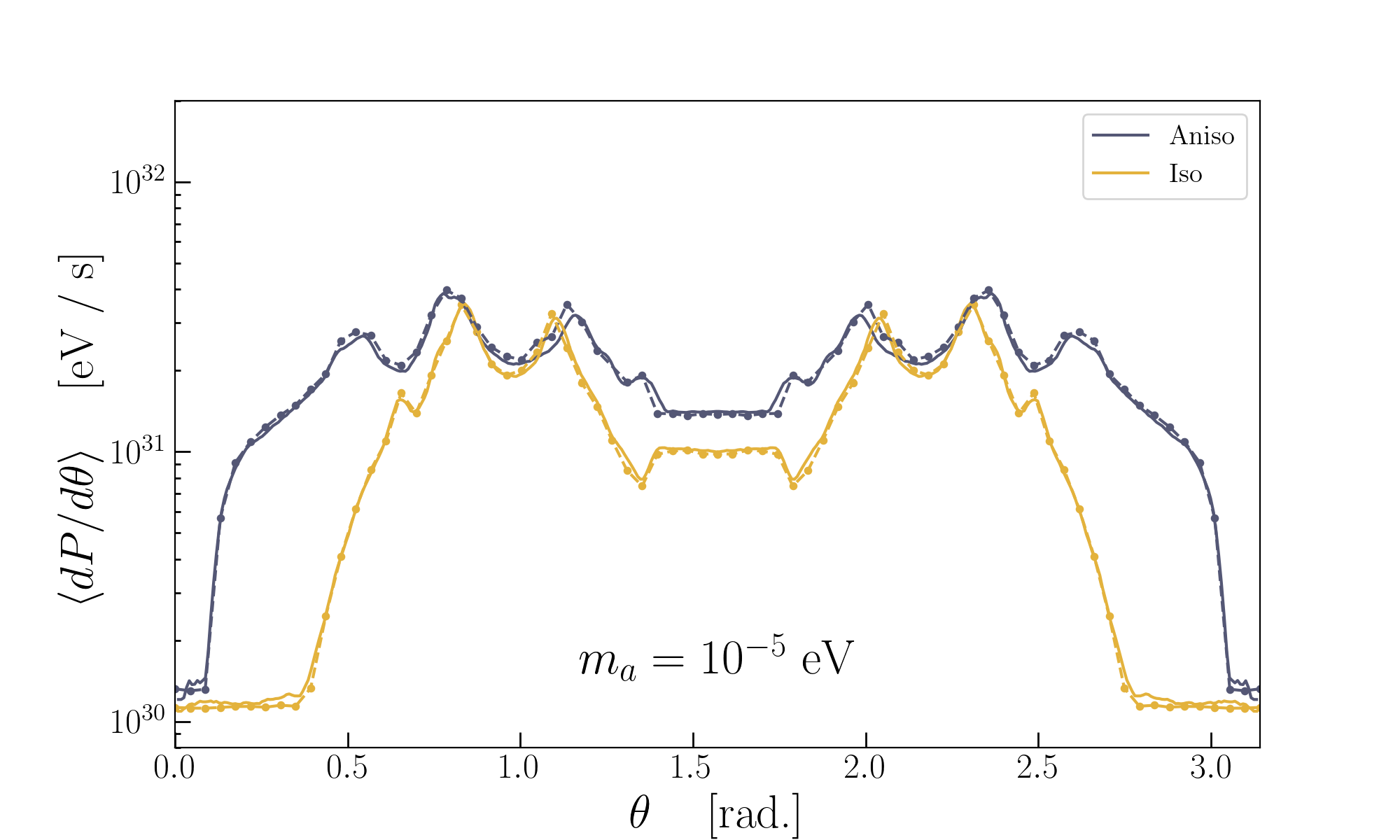}
\caption{\label{fig:jamiesam} \textbf{Equivalence of Forward and Backward Ray Tracing.} Period-averaged differential power as would be seen by an observer viewing at an angle $\theta$ with respect to the rotational axis, computed using forward (solid) and backward (dashed, points) ray tracing. Results are shown for the isotropic (`iso') and anisotropic ('aniso') dispersion relations  and an axion mass of $m_a = 10^{-5}$ eV. 
}
\end{figure}

The generalization of the conversion probability to the case of an isotropic plasma in curved spacetime is straightforward. The dispersion relation is given by
\begin{eqnarray}\label{eq:DispIsotropic}
   g^{\mu \nu} k_\mu k_\nu + \omega_p^2 = 0,
\end{eqnarray}
which gives
\begin{eqnarray}\label{eq:energyIso}
	E^{\rm iso \, ^2}_\gamma = \left| \textbf{k} \right|^2 + \omega_p^2.
\end{eqnarray}
From Eq.~\eqref{eq:Hypersurfaces}, this implies that family of surfaces $\Sigma_\textbf{k}$ collapses to a single emission surface $\Sigma$, on which
\begin{equation}
\omega_{\rm p}  = m_a,
\end{equation}
which is independent of $\textbf{k}$. In that case, the unit normal to surface $\Sigma$ is given by
\begin{eqnarray}
n^{\rm iso}_\mu =   \frac{\partial_\mu \omega_p^2}{\sqrt{\partial (\omega_p^2) \cdot \partial (\omega_p^2)} }
\end{eqnarray}
and the angle in Eq.~\eqref{eq:AngleDef} between this normal and the phase velocity, is given by
\begin{equation}\label{eq:CosIso}
	\cos \theta^{\rm iso}_\textbf{n} =
	\frac{k \cdot  \partial_\mu \omega_p^2}{
	\sqrt{[\partial (\omega_p^2) \cdot \partial (\omega_p^2)]\, [ k\cdot k]} }
\end{equation}
We also generalize the conversion probability in the isotropic case by modifying the flat space results according to
\begin{equation}
	\textbf{v}_a\cdot \nabla_\textbf{x} E^{\rm iso}_\gamma \rightarrow (k^\mu/E^{\rm iso}_\gamma) \cdot \partial_\mu E^{\rm iso}_\gamma
\end{equation}
where in the isotropic case we have
\begin{equation}
	\partial_\mu E^{\rm iso}_\gamma = \frac{\partial_\mu (\omega_p^2)}{2 E^{\rm iso}_\gamma} \, .
\end{equation}
Collectively, this yields a conversion probability given by
\begin{equation}\label{eq:ProbIsoCurved}
    P^{\rm iso}_{a \gamma } = \pi  g_{a \gamma \gamma}^2\left| \textbf{B}_{\rm ext} \right|^2 \sin ^2 \theta \frac{E_\gamma}{\left|k \cdot \partial (\omega^2_p) \right|} .
\end{equation}
If this is inserted into Eq.~\eqref{eq:PowerCurved}, the divergence occurring in the conversion probability Eq.~\eqref{eq:ProbIsoCurved} (arising when $k \cdot \partial \omega_p^2 \rightarrow 0$), is regulated by the cosine angle in the measure of Eq.~\eqref{eq:PowerCurved}, such that
\begin{align}
	\mathcal{P}^{\rm iso} & = \int d^3 \mathbf{k} \int d \Sigma \cos \theta_\mathbf{n} v_a E_\gamma^{\rm iso } P_{a \gamma} \, f_a \nonumber \\
	& = \int d^3 \mathbf{k} \int d\theta d\varphi \sqrt{|h_\mathbf{k}|} \, \, E_\gamma^{\rm iso } \, \frac{ \pi g_{a \gamma \gamma}^2 \left| \mathbf{B}_{\rm ext}\right|^2 }{\left| \partial (\omega_p^2)\right|}.
\end{align}
where $\left| \partial (\omega_p^2)\right| = \sqrt{\partial(\omega_p^2) \cdot \partial (\omega_p^2)}$ is the modulus of $\partial_\mu (\omega_p^2)$. This is manifestly convergent.

Generalising Eq.~\eqref{eq:ProbIsoCurved} to the anisotropic case in curved spacetime is non-trivial, as this involves formulating kinetic theory for photons in curved spacetime~\cite{Acuna-Cardenas:2021nkj} -- note that this is similar to what is done for the case of scalars in Ref.~\cite{Hohenegger:2008zk}. Such a generalization is clearly important,
but lies  beyond the scope of the present work. Instead, for the purpose of making progress, we employ a phenomenological generalization of the conversion probability similar to what is presented above, taking as before the substitution
\begin{equation}
	\textbf{v}_a\cdot \nabla_\textbf{x} \rightarrow (k^\mu/E_\gamma) \cdot \partial_\mu E_\gamma
\end{equation}
to arrive at an ansatz for the \textit{anisotropic} conversion probability in curved spacetime of the form
\begin{eqnarray}\label{eq:CPAnisoCurved}
    P^{\rm aniso}_{a \gamma } = \frac{\pi}{2}\, \frac{ g_{a \gamma \gamma}^2\left| \textbf{B}_{\rm ext} \right|^2E_\gamma ^4 \sin ^2 \theta }{\cos ^2 \theta \,
   \omega_p^2 \, \left(\omega_p^2-2 E_\gamma ^2\right)+E_\gamma ^4} \,    \frac{E_\gamma}{\left| k \cdot \partial E_\gamma  \right|} .
\end{eqnarray}
Similarly, we generalize the angle in curved space by taking
\begin{eqnarray}\label{eq:AngleDefAniso}
	\cos \theta_\textbf{n} = \frac{k \cdot \partial E_\gamma}{|k|  \, |\partial E_\gamma|}
\end{eqnarray}
which guarantees that the differential power computed using the forward propagation approach is convergent.
However, since this is only a partial generalization to curved spacetime, divergences are not canceled in the  backward ray tracing approach (see Appendix \ref{sec:APPCurved} for a more detailed discussion of the origin of these divergences). In order to regulate these divergences, we introduce an IR cutoff in the backward ray tracing approach on the effective conversion length, defined by $L_c \equiv \hat{k} \cdot \partial E_\gamma$, which serves to exclude any contribution with $L_c > 1 {\rm km}$\footnote{We have verified that adjusting this threshold by a factor of 2 leads to negligible changes in asymptotic power.}.
This should not be confused with any type of formal regulatory procedure; rather it is a means to excise points which lead to divergences that would otherwise be regulated with  a formal generalization of the conversion probability to curved space time.

\begin{figure*}
\includegraphics[width=0.49\textwidth]{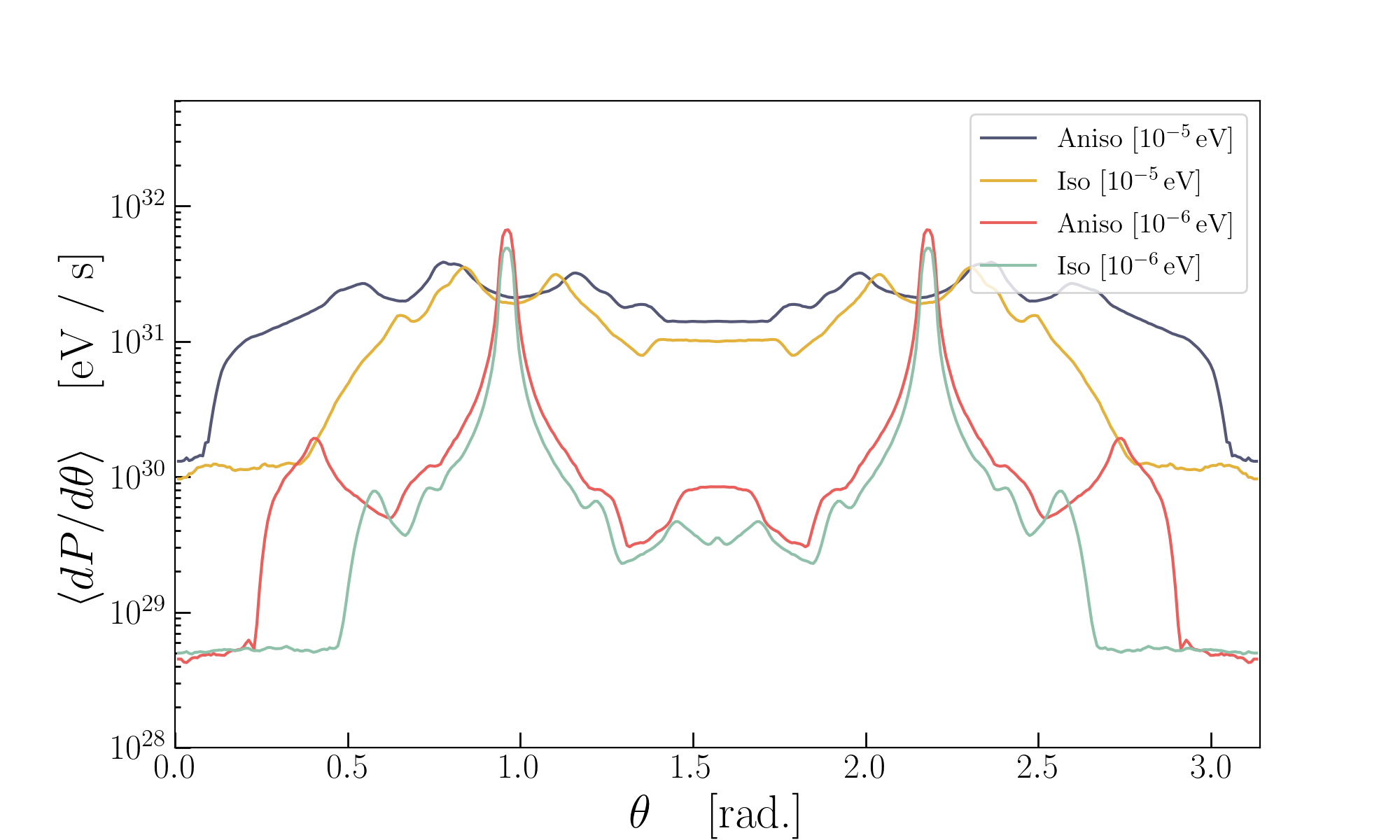}
\includegraphics[width=0.49\textwidth]{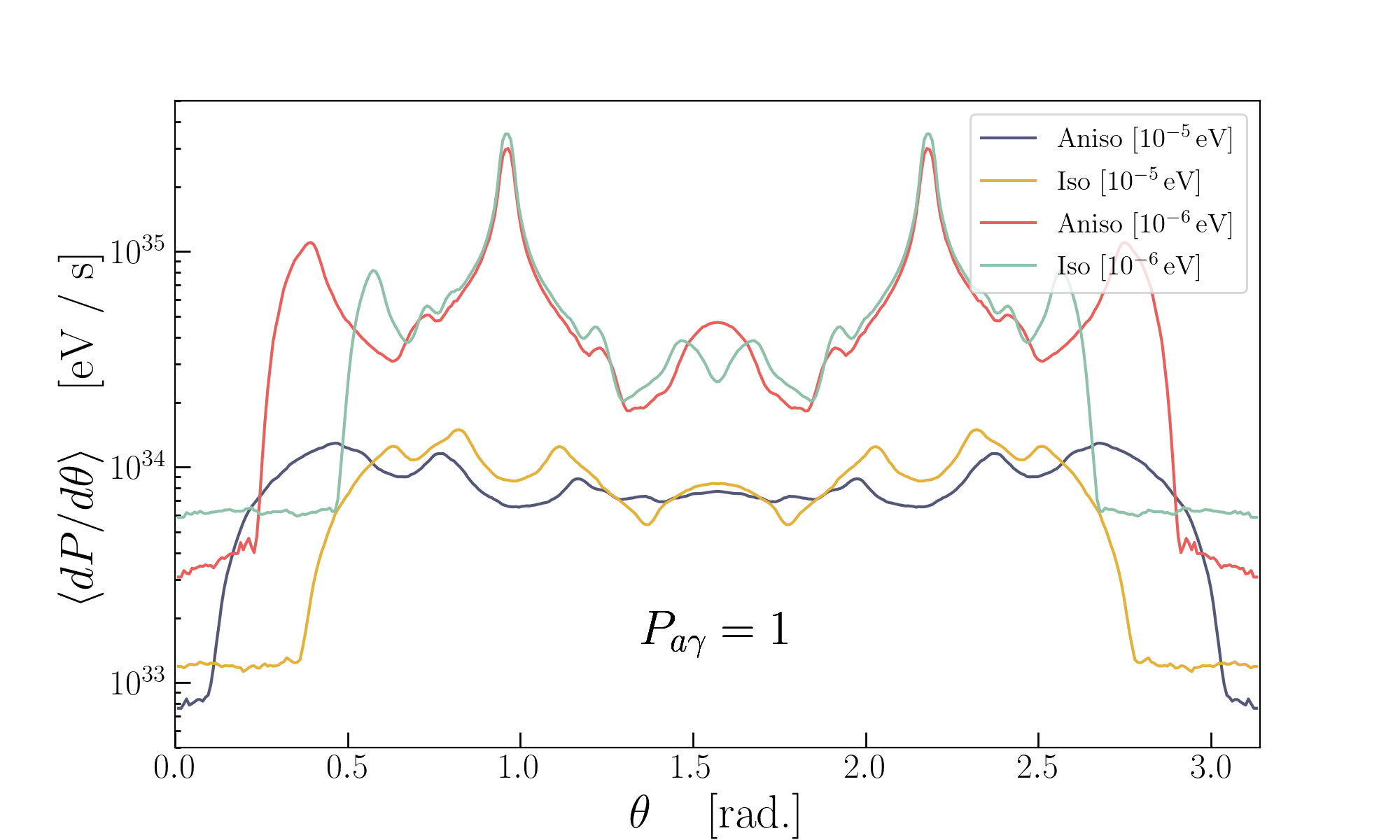}
\caption{\label{fig:flatiso} \textbf{Effects of Plasma Anisotropy.} Period-averaged differential power as would be seen by an observer viewing at an angle $\theta$ with respect to the rotational axis.  Results are shown for two choices of the axion mass ($m_a = 10^{-5}$ and $10^{-6}$ eV).  The isotropic (`iso') case combines a dispersion relation  $g^{\mu \nu}k_\mu k_\nu + \omega_p^2 = 0$ and conversion probability $P^{\rm iso}_{a \gamma}$ of Eq.~\eqref{eq:ProbIsoCurved}.  The anisotropic (`aniso') uses the dispersion relation of Eq.~\eqref{eq:AnisotropicDisp} and the conversion probability $P_{a \gamma}^{\rm aniso}$ in Eq.~\eqref{eq:CPAnisoCurved}. All results include gravity. Plots are generated using the fiducial parameters of the main text.  }
\end{figure*}

\section{Axion Dark Matter Detection with Neutron Stars} \label{sec:NS}

Having outlined the generalized ray tracing procedure, we can now return to the problem at hand -- namely, the application of these approaches to radio searches for axion dark matter near neutron stars. We choose to focus in particular on spectral lines arising from a smooth background distribution of axion dark matter, which are among the most well-studied indirect probes of axions in these environments \cite{Pshirkov:2007st,Hook:2018iia, Huang:2017egl, Leroy:2019ghm, Safdi:2018oeu, Battye:2019aco, Foster:2020pgt, Prabhu:2020yif, Foster:2022fxn, Witte:2021arp,millar2021axionphotonUPDATED, Battye:2021yue,Battye:2023oac}. One should bear in mind, however, that these tools are more broadly applicable to similar searches, such as those looking for transient lines from the encounters of miniclusters and axion stars with neutron stars~\cite{Buckley:2020fmh,Edwards:2020afl,Bai:2021nrs,Nurmi:2021xds,Witte:2022cjj}, and broadband radio searches generated from a locally sourced population of axions~\cite{Prabhu:2021zve,Noordhuis:2022ljw,Noordhuis:2023wid}.

We begin by establishing a set of fiducial parameters which define our baseline model, given by an axion mass of $m_a = 10^{-5} \, {\rm eV}$, an axion-photon coupling $g_{a\gamma\gamma} = 10^{-12} \, {\rm GeV^{-1}}$, a neutron star mass of $M_{\rm NS} = 1 M_{\odot}$, a neutron star radius of $r_{\rm NS} = 10$ km, a surface dipolar magnetic field of $B_0 = 10^{14}$ G, a neutron star rotational frequency of $\Omega_{\rm NS} = 1 $Hz, and a misalignment angle $\theta_m = 0$ radians. In what follows we normalize the axion distribution to asymptotic number density $n_{a,\infty} = 0.45 \times m_a^{-1} \, {\rm GeV \, cm^{-3}}$. Deviations from these parameters below are always explicitly stated.

Each of the examples in this section is illustrated assuming a perfectly dipolar magnetic field and a GJ charge density, defined by a $e^\pm$ charge density
\begin{eqnarray}
n_{\rm GJ} & \simeq & \frac{2 \vec{\Omega} \cdot \vec{B}}{e}  \, \nonumber \\
& \simeq & \frac{\Omega \, B_0}{e} \left(\frac{r_{\rm NS}}{r} \right)^3 \, \left[3 \cos\theta \hat{m} \cdot \hat{r} - \cos\theta_m \right] \label{eq:GJDensity}\, ,
\end{eqnarray}
where $\theta_m$ is the misalignment angle between the rotational and magnetic axis, and the time dependence of the plasma due to rotation has been embedded in the term $\hat{m} \cdot \hat{r} = \cos\theta_m \cos\theta + \sin\theta_m \sin\theta \cos(\Omega t)$, where $\theta$ is a polar angle given by taking rotational axis as the north pole. Throughout this paper, we choose coordinates such that $\theta$ also gives the angle between line of sight to an observer and the rotational axis. The GJ charge distribution is expected to be a reasonable approximation across most of the closed field lines of standard pulsars, and near the surfaces of dead neutron stars~\cite{Spitkovsky:2002wg,petri2002global,Safdi:2018oeu}.
We revisit this assumption when applying our results to the galactic center magnetar, as the charge densities near such objects are expected to be highly enhanced with respect to the GJ model~\cite{Thompson:2020hwt}.

Before illustrating how each assumption and free parameter impacts the projected radio signal, we begin by illustrating the agreement between forward and backward ray tracing, using the formalism developed in the preceding sections. In Fig.~\ref{fig:jamiesam} we show the differential power (averaged over the rotational period of the neutron star) generated from resonant axion-photon transitions, and computed using either an anisotropic, or an isotropic, dispersion relation in curved spacetime. We show results for an axion mass of $10^{-5}$ eV, but have also confirmed agreement at other masses.
These numerical results confirm the equivalence of forward and backward ray tracing, in accordance with the theory laid out in Sec.~\ref{sec:raytrace}. This demonstrates explicitly the equivalence of sampling the collision integral and explicitly evaluating the asymptotic flux via back-tracing, with each scenario manifested in the right- and left-hand side of Eq.~\eqref{eq:Continuity}, respectively.

Before continuing, it is worth highlighting that achieving such a high level of agreement requires not only a unified formalism (see Sec.~\ref{sec:raytrace}), but also high-precision numerics, with the results being strongly sensitive to a variety of different factors, such as: the initial conditions of the photon in the backward ray tracing procedure, the value of fundamental constants (we find maintaining consistency across many decimal places is typically required), the precision of the ODE solver, etc. We now go on to  illustrate how varying the size of different physical phenomena discussed in this and proceeding sections affect observational signatures from axion dark matter conversion in neutron stars.

\subsection{Physics of ray tracing}

\begin{figure*}
\includegraphics[width=0.49\textwidth]{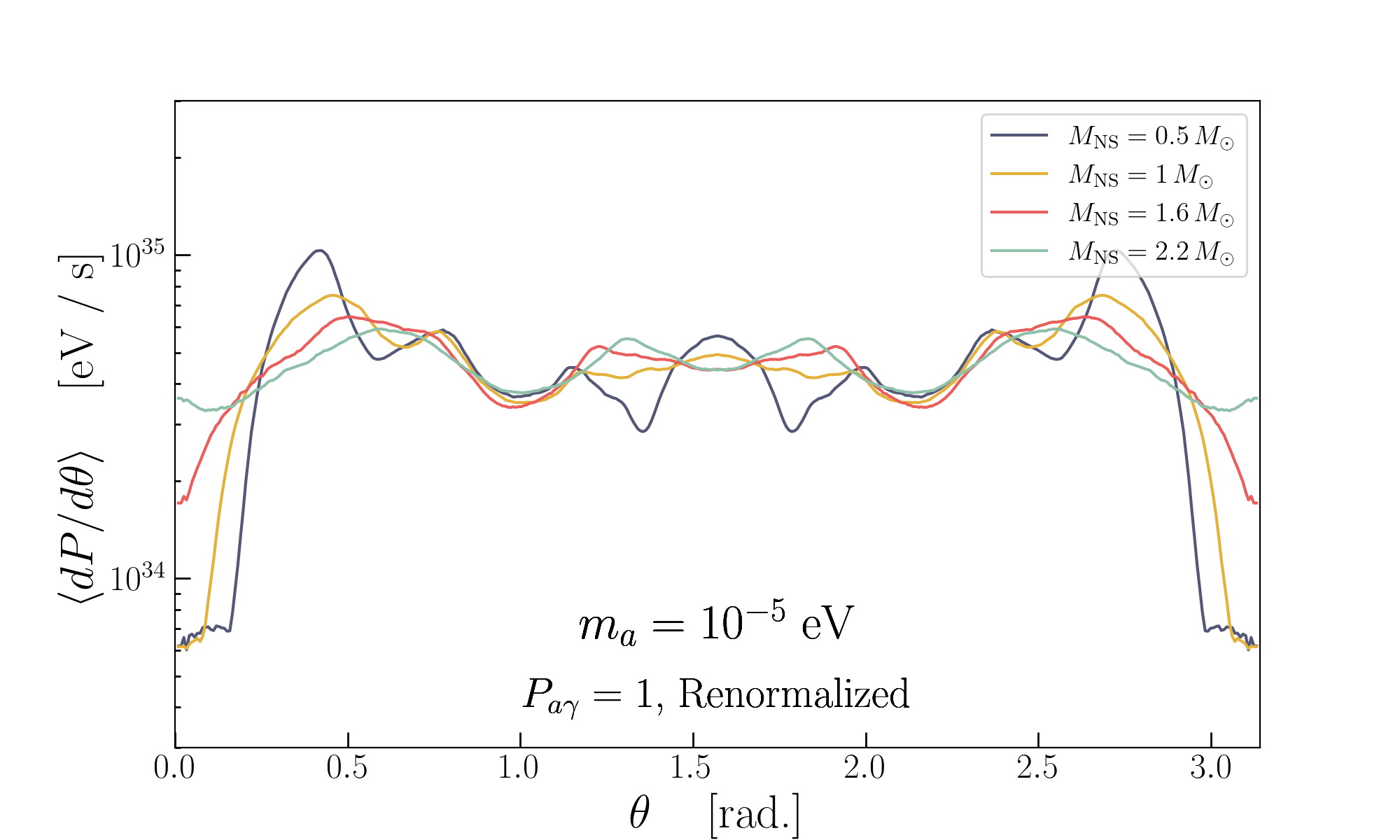}
\includegraphics[width=0.49\textwidth]{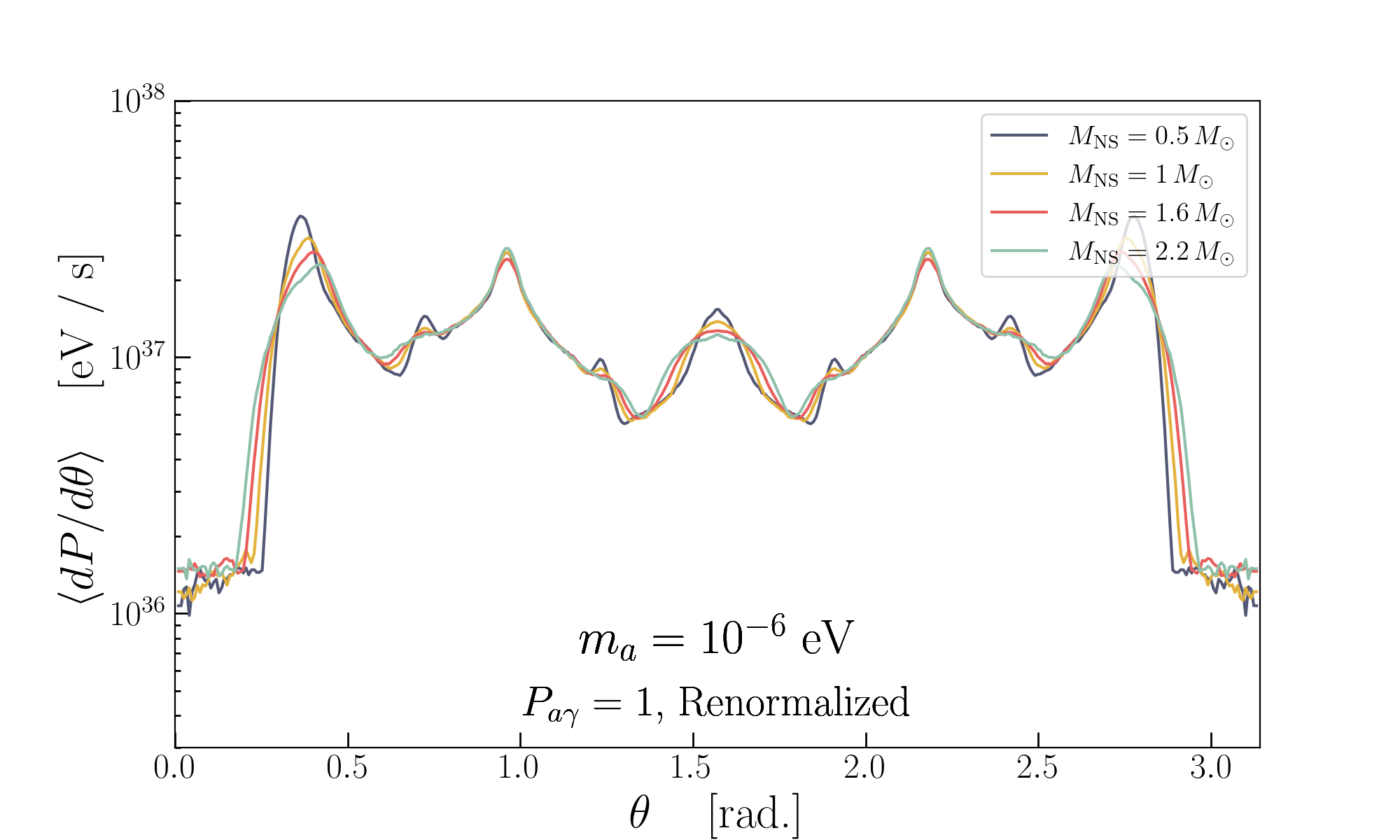}
\caption{\label{fig:MassNS_Renorm} \textbf{Neutron Star Mass (Renormalized).} Same as Fig.~\ref{fig:MassNS}, but setting the axion-photon conversion probability to 1 for all rays, and re-normalizing the weights of each ray by the factor $\mathcal{R}_i$ given in Eq.~\ref{eq:reweight} (this re-normalization ensures the sky-integrated power is fixed across all neutron star masses). Result is shown for an axion mass of $m_a = 10^{-5}$ eV (left) and $10^{-6}$ eV (right).
}
\end{figure*}

The goal of this subsection is to illustrate the importance of the various physical effects within our models, which until now, have not yet been self consistently incorporated. This includes the combined impact of plasma anisotropy and curved space on photon propagation, the effect of varying the neutron star mass and radius, the importance of including multiply reflected photons, and the impact of simplifications to the asymptotic velocity distribution of axions prior to in-fall. In Appendix \ref{sec:APPcomparison} we also examine the extent to which gravity can be encoded in the initial conditions of the photons, and the importance of imposing the proper kinematic matching condition\footnote{We reserve these comparisons for the Appendix as they are not fully self-consistent, and as such both their implementation and interpretation is something of a subtlety that could cause confusion for less familiar readers. }. In each case, we analyze the impact of including and/or neglecting an effect by computing the rotation-period averaged differential power $\left< d \mathcal{P}/d\theta \right>$ as a function of the viewing angle, $\theta$. We emphasize that it is not always straightforward to isolate individual effects, since a self-consistent treatment usually involves not just one, but many modifications. As an example, one can consider that gravity plays a role not only in modifying the trajectories of individual rays, but also enters the photon initial conditions (which in turn enter the conversion probability, producing an effect both on the weighting of photons and their propagation) and axion number density. We clarify below when isolating an effect breaks self-consistency.

In the next two subsections, we begin by addressing the question of the relative importance of treating the anisotropy of the plasma and curve spacetime, the former having been treated in~\cite{Witte:2021arp} and latter in~\cite{Battye:2021xvt}.

\begin{figure*}
\includegraphics[width=0.49\textwidth]{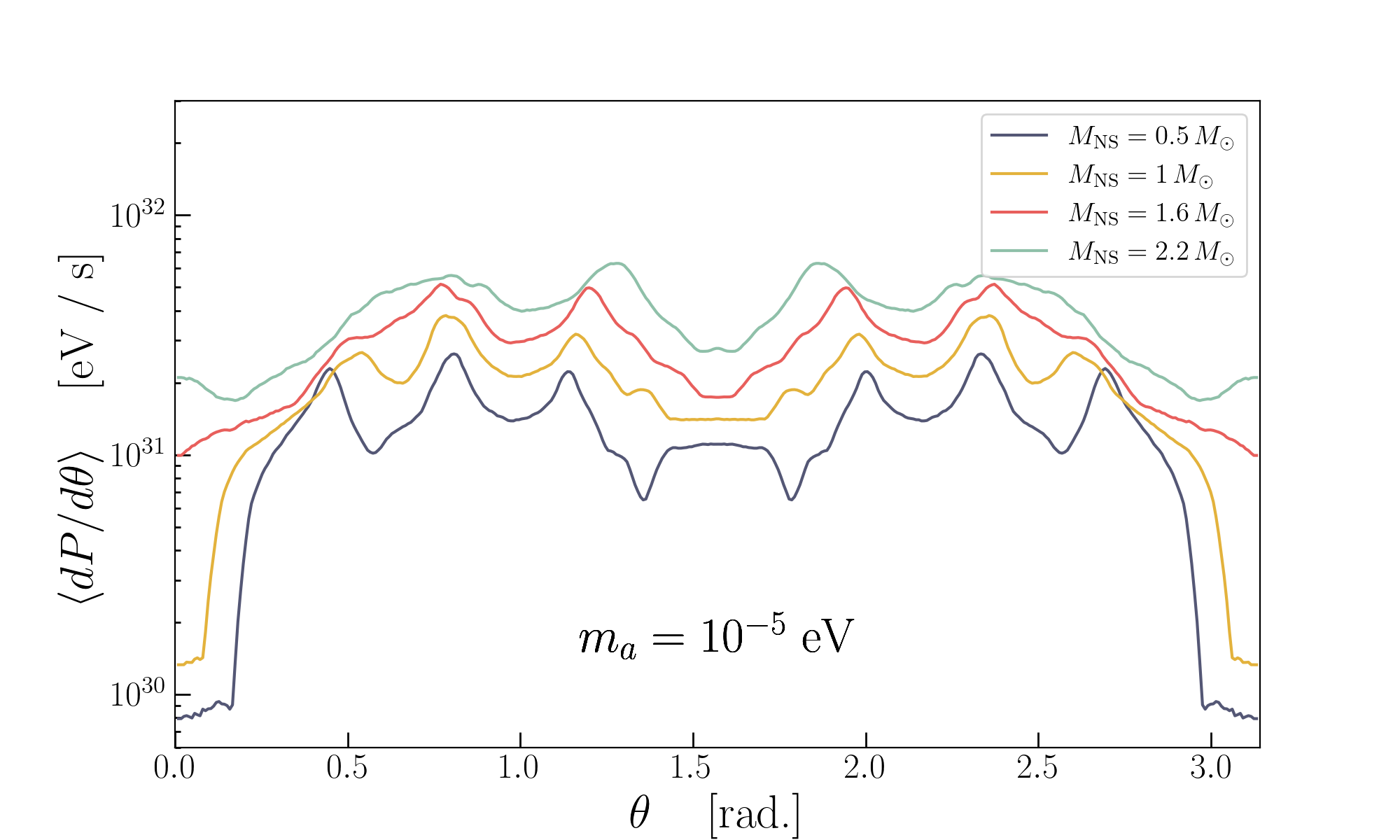}
\includegraphics[width=0.49\textwidth]{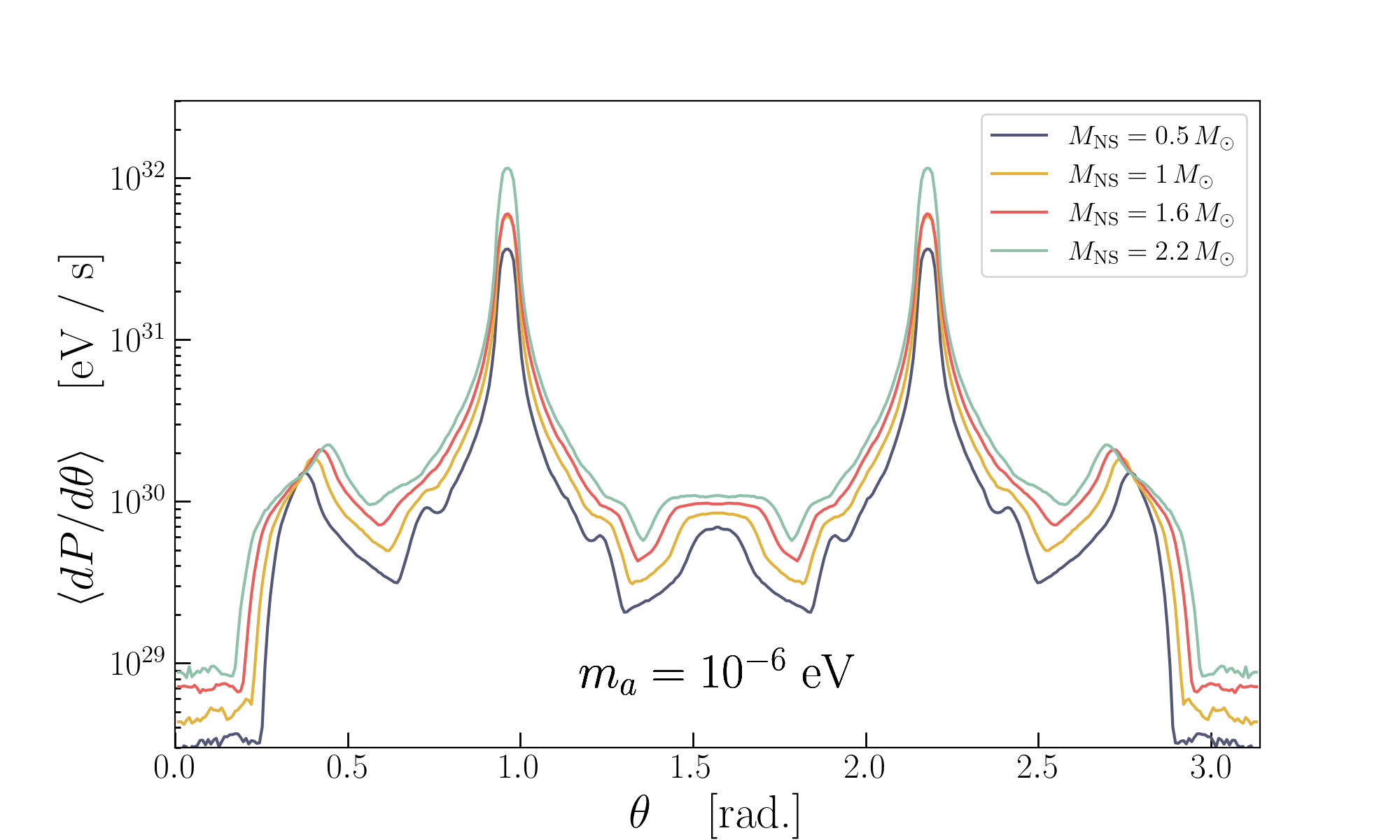}
\caption{\label{fig:MassNS} \textbf{Neutron Star Mass.} Same as Fig.~\ref{fig:flatiso}, but showing the impact of varying the neutron star mass from $1 \, M_\odot$ to $2.2 \, M_\odot$.  Result is shown for an axion mass of $m_a = 10^{-5}$ eV (left) and $10^{-6}$ eV (right). Bottom panel in each case shows the relative difference (in percent) with respect to the fiducial neutron star mass of $1 \, M_\odot$.
}
\end{figure*}

\subsubsection{Plasma Anisotropies}

We now examine differential power for an axion mass of $10^{-5}$ eV and $10^{-6}$ eV, assuming either an isotropic or anisotropic plasma, which correspond to the dispersion relations of Eqs.~\eqref{eq:DispIsotropic} and \eqref{eq:AnisotropicDisp}, respectively.  
In each case, we also adopt a conversion probability that is self-consistent with the choice of dispersion relation; namely, we use $P_{a \gamma}^{\rm iso}$ (Eq.~\eqref{eq:ProbIsoCurved}) for the isotropic plasma, and $P_{a \gamma}^{\rm aniso}$ (Eq.~\eqref{eq:CPAnisoCurved}) for the anisotropic plasma. Modifying the photon dispersion relation only redistributes outgoing photons in phase space, keeping modifying how power is distributed on the sky while maintaining a fixed sky-integrate power (assuming photon absorption can be neglected). Meanwhile, changing the conversion probability between the isotropic and anisotropic cases affects the overall normalization of the power, in accordance with Eq.~\eqref{eq:ForwardTracingEquation1}. In particular, this leads to a larger overall power output for an anisotropic plasma. For comparison, we also plot a comparison between the anisotropic and isotropic scenarios, but setting the conversion probability in both cases to 1 for all rays -- this allows us to isolate the impact of plasma anisotropy on the evolution of photon trajectories. The effect of gravity is included in all cases.

The results for isotropic and anisotropic plasmas are shown in Fig.~\ref{fig:flatiso} (the left panel includes the conversion probability, while in the right panel it is set to unity). We see that for the parameters chosen, the axion mass plays the dominant role in changing both the morphology and the normalization of the power profile, with the plasma anisotropy driving a more subtle, although non-negligible, change of shifting power away from the poles and equatorial plane -- this latter effect arises from the $\theta_B$ dependence in the photon dispersion relation. At larger masses ($m_a= 10^{-5}$ eV), we see that differential power varies by an order of magnitude across $\theta \in (0, \pi)$, while the effect of assuming an isotropic plasma can induce variations (for fixed values of $\theta$) at the order of magnitude level, although it is worth noting that the average difference is not so pronounced. For an axion mass of $10^{-6}$ eV (where the conversion surface extends to larger characteristic radii), the differential power across the sky can vary by more than three orders of magnitude, with the effect of plasma anisotropy introducing as much as a one order of magnitude shift in the power (although only a certain viewing angles).

A range of approximations \cite{Hook:2018iia,ref:NS-Japan,Leroy:2019ghm,Witte:2021arp,Battye:2023oac,Foster:2020pgt,Foster:2022fxn} have been used throughout the literature to model plasma physics in the context of axion searches, both at the level of dispersion relations and production itself. In this work, we are now in a position to collate these various approximations and make some assessment of their relative importance from a observational point of view. More explicitly, given the results reported in Fig.~\ref{fig:flatiso}, the question  arises as to the significance of the differences between simplified isotropic scenarios and more complete anisotropic plasma descriptions, the latter combining anisotropic effects  both in dispersion relations and conversion probabilities.

One should bear in mind that the results of Fig.~\ref{fig:flatiso} concern aligned rotators ($\theta_m = 0$), whilst in general neutron stars are non-aligned with $\theta_{\rm m} \neq 0$. In the case of a frequency domain analysis (in which one uses time-integrated observations), the primary effect of inducing a misaligned rotation axis is to isotropize the period-averaged flux (see \eg the differential power curves in~\cite{Witte:2021arp}).

\subsubsection{Curved Spacetime Plasma Effects}

Having discussed the effect of plasma anisotropy, we turn now to the role of curved spacetime effects. We begin by focusing on the role of gravity in altering the photon dispersion relation (and hence ray-propagation)~\cite{McDonald:2001vt}, and then return to the impact of gravity on the local intensity emitted at the conversion surface.

In order to understand the effect of gravity we begin by running the ray tracing analysis using a variety of different neutron star masses ranging from $M_{\rm NS} = 0.5 \, M_\odot$ to $M_{\rm NS} = 2.2 \, M_\odot$. We begin by isolating the effect of gravity on the propagation of photons by re-scaling the photon weights in such a way that the sky-integrated asymptotic power remains fixed. This is done by setting the conversion probability of each ray to one, and re-scaling each photon by a factor
\begin{eqnarray}\label{eq:reweight}
	\mathcal{R}_i \equiv \frac{1 \, {\rm cm^{-3}}}{n_a(r_i)} \, \frac{m_a}{|\vec{k}_a(r_i)|} \, \frac{1}{\sqrt{1-r_s / r_i}} \, ,
\end{eqnarray}
where $r_i$ is the initial radius of the photon at the resonance. Note that the choice of normalization of the axion number density and photon  momentum are arbitrary, and thus one should not attempt to interpret the normalization of the differential power as being physical. The result of this procedure is shown for two choices of axion mass in Fig.~\ref{fig:MassNS_Renorm}. For $m_a = 10^{-5}$ eV (\ie when the conversion surface is close to the neutron star), the effect is most pronounced near the magnetic poles, leading to a variation in the differential power up to a factor of $\sim 5$ -- at other viewing angles, however, the effect is typically no more than factor of two. As expected, the effect of increasing the neutron star mass leads to an isotropization of the radio flux~\cite{Battye:2021xvt}. For smaller axion masses, the conversion surface shifts away from the neutron star and the effect of gravity on the propagation of rays is suppressed; specifically, for the case of $m_a = 10^{-6}$ eV, we find gravity only modifies ray propagation at the $\mathcal{O}(10\%)$ level. In the Appendix, we attempt to understand the extent to which previous approximations adopted in~\cite{Witte:2021arp} are valid; in particular, we show that embedding the effect of gravity into the initial conditions, but neglecting gravity during the propagation, tends to induce negligibly small errors in the differential power (we caution, however, that this is an empirical result, and is not guaranteed to hold in all contexts).

We now return to analyzing the full effect of gravity, which enters not only the propagation of rays but also the local power emitted from the resonant conversion surface. We plot in Fig.~\ref{fig:MassNS} the impact of increasing the neutron star mass from $1 \, M_\odot$ to the maximally allowed value of $\sim 2.2 \, M_\odot$ for two choices of axion masses (unlike Fig.~\ref{fig:MassNS_Renorm}, here we maintain the conversion probability and the appropriate weights of the rays).  Fig.~\ref{fig:MassNS} illustrates that the neutron star mass plays a non-negligible role in the prediction of the radio flux, and should likely be included in future modeling.  

\begin{figure}[t!]
\includegraphics[width=0.49\textwidth]{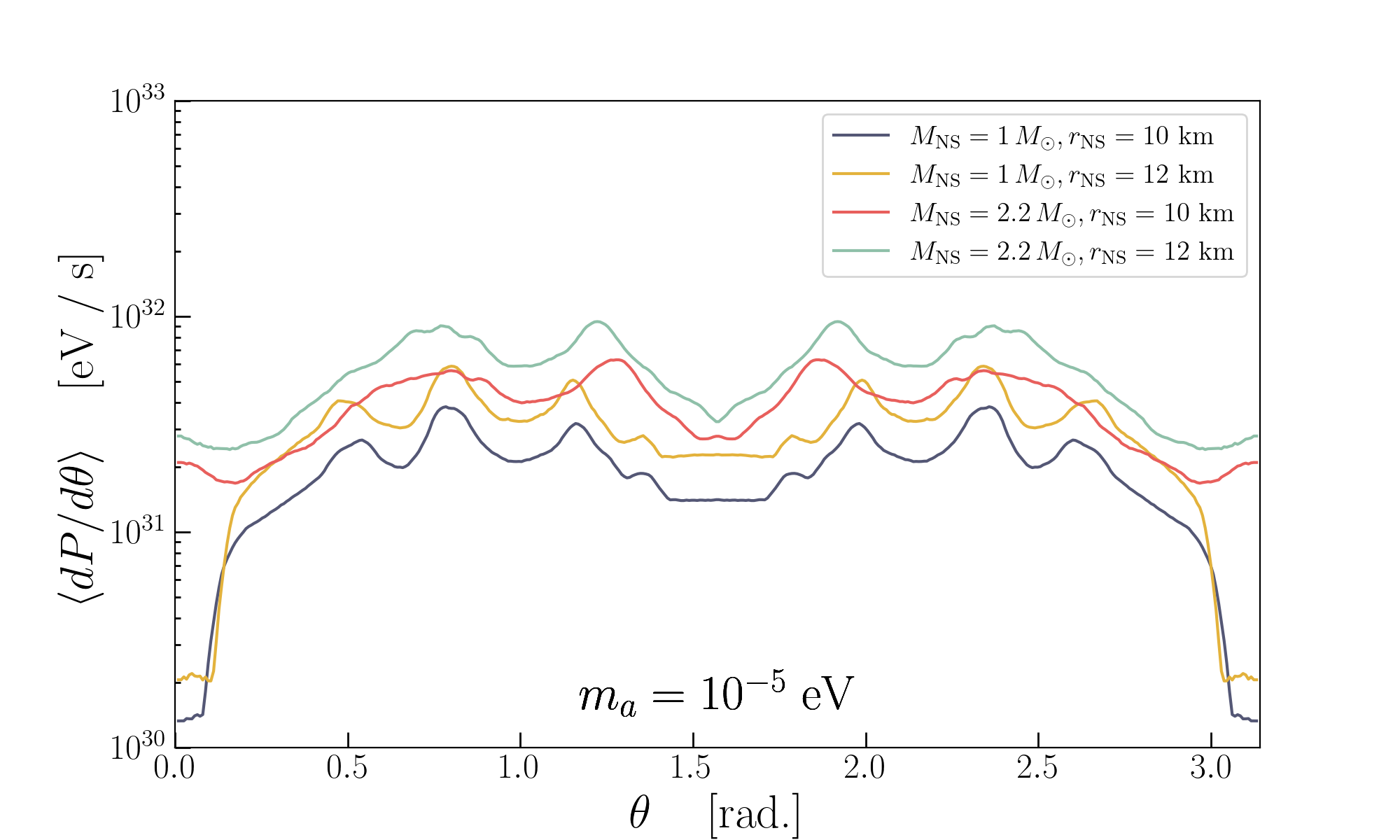}
\caption{\label{fig:RadiusNS}\textbf{Neutron Star Radius.} Same as left panel of Fig.~\ref{fig:flatiso} but varying the neutron star radius and neutron star mass between $r_{\rm NS} \in [10, 12]$ km and $M_{\rm NS} \in [1, 2.2] \, M_\odot$. As before we show in the bottom panel the relative difference (in percentage) with respect to the fiducial models (taken to be those with $r_{\rm NS} = 10$ km). }
\end{figure}

In Fig.~\ref{fig:RadiusNS} we illustrate the impact of varying the neutron star radius, assuming a neutron star mass of either $1 \, M_\odot$ or $2.2 \, M_\odot$. Here, we see that the characteristic neutron star radius adopted in previous works tends to lead to an underestimation of the radio flux by a factor of $\sim 2$ -- this effect is nearly uniform over the sky, and arises primarily from the fact that the increase of radius (at fixed surface magnetic field strength) tends to increase the net surface area over which resonant axion-photon transitions take place.

The initial ray tracing performed in Refs.~\cite{Witte:2021arp} and ~\cite{Battye:2021xvt} had observed notable deviations in the inferred time profiles induced by resonant axion-photon mixing (note that time variation is directly related to the variation of the differential power with viewing angle, $\theta$). Ref.~\cite{Battye:2021xvt} had attributed this difference to the effect of gravity -- it was argued that gravity tends to isotropize the signal, washing out the strong variation observed in the time profiles of ~\cite{Witte:2021arp}. Here, we show that this conclusion is not correct; instead, the variation observed in Ref.~\cite{Witte:2021arp} was larger due to three contributing factors: $(i)$ ~\cite{Witte:2021arp} focused on lower mass axions, which have larger time variation due to the larger characteristic conversion surface, $(ii)$ Ref.~\cite{Battye:2021xvt} had not included multiply reflected photons, which tends to increase the time variation (see Fig.~\ref{fig:MultiReflections}, and the following subsection), and $(iii)$ the anisotropy of the plasma at low masses can further enhance the variation in the time-domain.

\subsubsection{Multiply Reflected Photons}\label{sec:multipleReflection}

The backward ray tracing method \cite{Battye:2021xvt} is essentially a standard approach to the problem of radiative transfer \cite{Befki1966}. Photon production is calculated via an integral along the photon worldline, as depicted in Eq.~\eqref{eq:BoltzWorldine}. As the photon worldline is back-propagated, it can encounter multiple sites of photon production from axions. This happens wherever axion and photon dispersion relations become degenerate,  corresponding to a family of discrete values $\lambda_i$ of the worldline parameter at which photons are produced.  In the original work \cite{Battye:2021xvt} we only considered the first level crossing\footnote{Strictly speaking, we multiplied $f_\gamma(\lambda_1)$ by a factor 2 (as in \cite{Leroy:2019ghm}) to model the effect of photons which are produced by axions moving in the opposite direction to the photon.} and $\lambda = \lambda_1$. In the present work, we include contributions to the asymptotic value of the photon distribution $f_\gamma(\lambda \rightarrow \infty )$ from all production sites along the photon worldline, so that  $f_\gamma(\lambda \rightarrow \infty ) = \sum_i f_\gamma(\lambda_i)$.  This effect is especially pronounced within the throats of the magnetosphere, where the worldines of photons experience multiple reflections, causing them to repeatedly encounter level-crossings.

\begin{figure}
	\centering
	\includegraphics[width=0.49\textwidth]{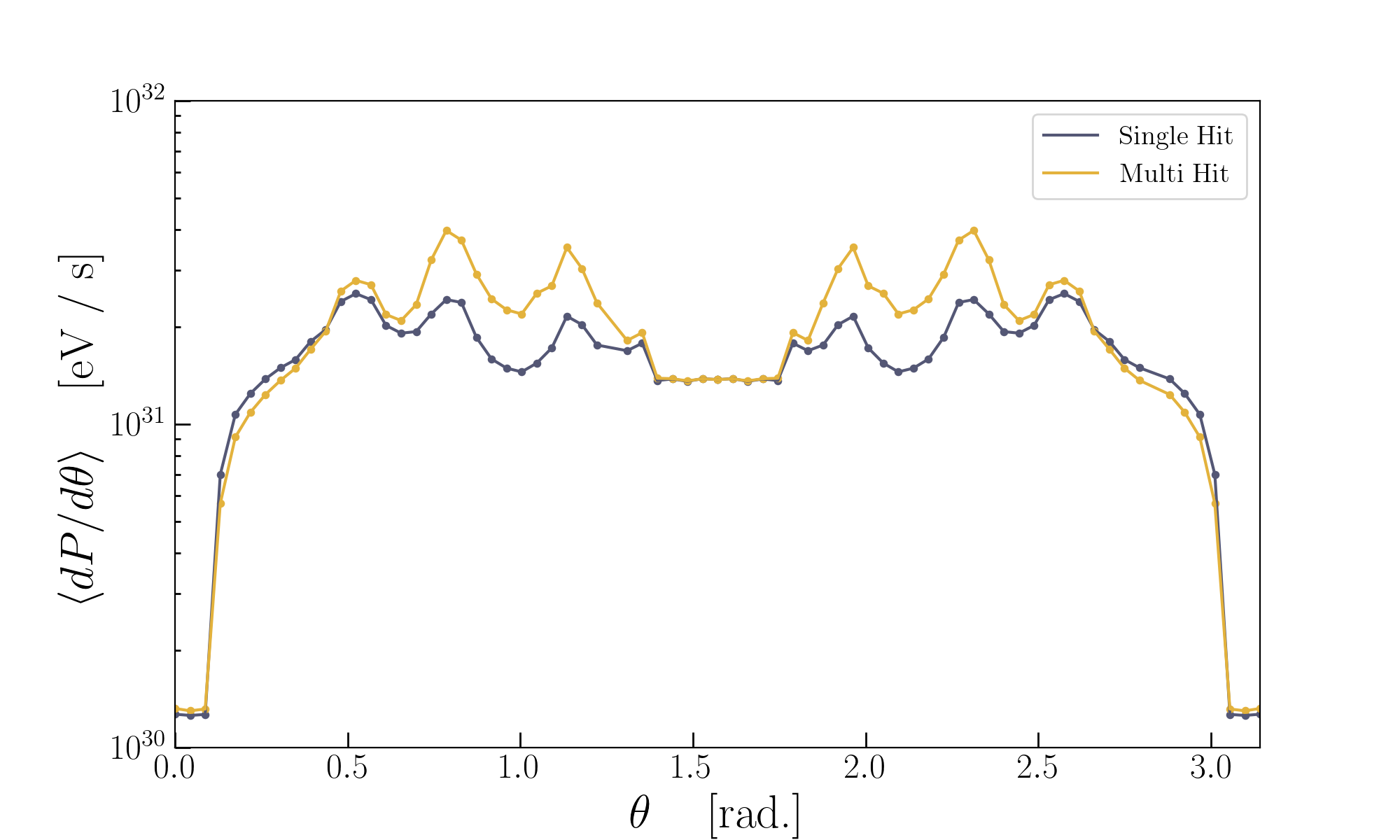}
	\caption{\textbf{Multiple Photon Production.} The impact of including multiple resonant photon productions along the photon worldline in backward ray tracing. For $m_a = 10 \mu{\rm eV}$ and the same fiducial NS parameters used in previous plots. }
	\label{fig:MultiReflections}
\end{figure}

In Fig.~\ref{fig:MultiReflections}, we illustrate the impact of including these multiply reflected photons for two different axion masses. Here, one can see that accounting for multiply reflected photons tends to induce a factor of $\sim 2$ enhancement in the differential power for viewing angles that are roughly aligned with the throats of the magnetosphere. This effect is enhanced for smaller axion masses, where the throats become more prominent.

\subsubsection{Asymptotic Velocity Distribution}

\begin{figure}
	\centering
	\includegraphics[width=0.49\textwidth]{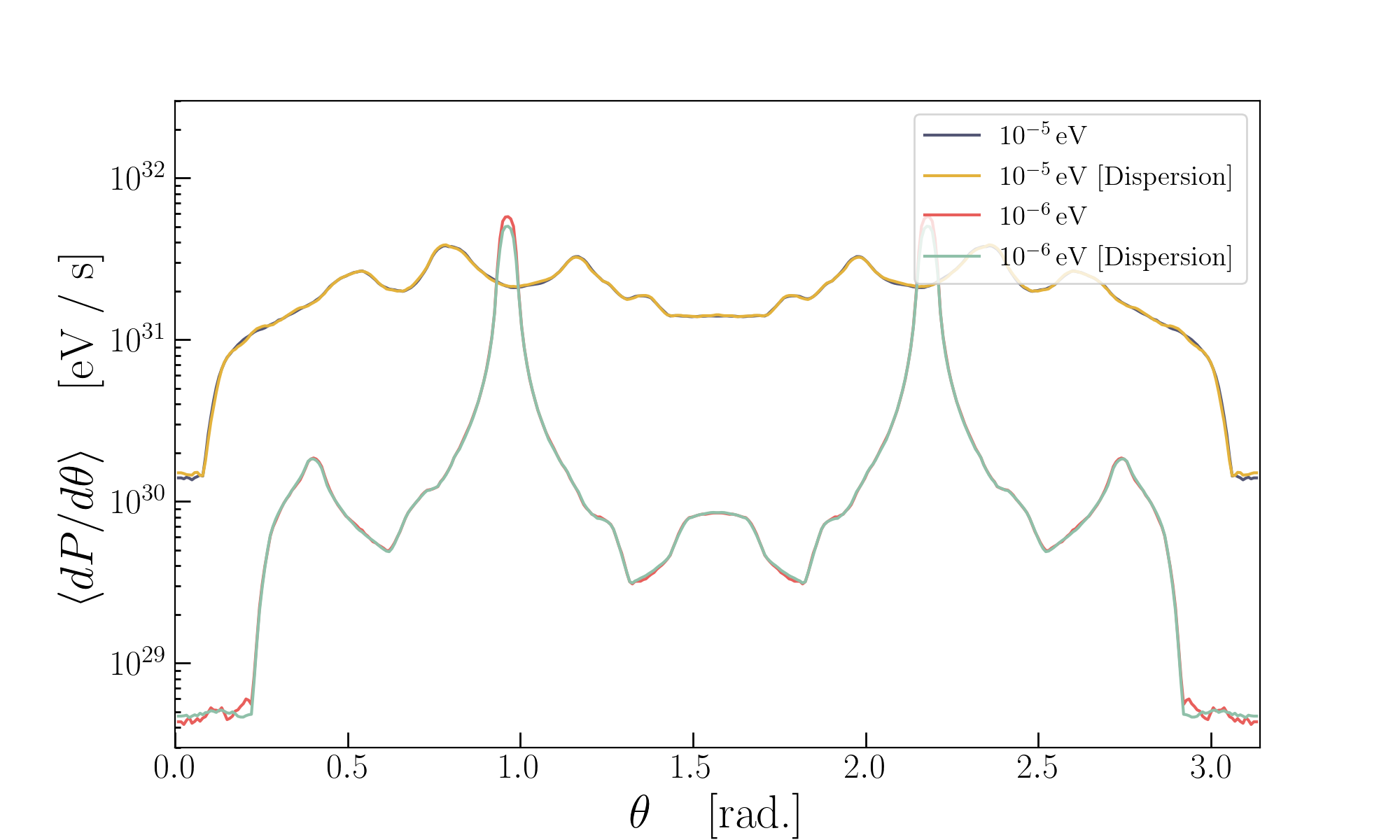}
	\caption{\textbf{Finite Line Width Effects.} The impact of including the velocity dispersion (as opposed to adopting a delta function on the asymptotic speed $|\vec{v}_\infty| = 220$ km /s for two choices of the axion mass. }
	\label{fig:vel_disp}
\end{figure}

The examples provided thus far have assumed for simplicity that the asymptotic axion speed distribution can be treated as a delta function fixed to $|\vec{v}| \sim 220$ km/s, corresponding to a monochromatic photon signal. This is the assumption adopted in~\cite{Battye:2019aco,Battye:2021xvt,Battye:2023oac}, while~\cite{Witte:2021arp,Foster:2022fxn}
 treated the full Maxwellian distribution. In Fig.~\ref{fig:vel_disp}, we show that for both axion masses of interest, the simplification of neglecting the width of the asymptotic energy distribution tends to induce negligible variations in the inferred power. Importantly, however, this statement is only valid when the neutron star is assumed to be at rest with respect to the galaxy, and may not hold for strongly boosted neutron stars.

\section{Galactic Center Magnetar}
The Galactic Center magnetar SGR J1745–2900 has an inferred dipolar field strength of $1.6\times10^{14} $G, a rotational period $P \sim 3.76$s, and has a two-dimensional projected distance from the Galactic Center of $\sim 0.17$pc~\cite{Mori:2013yda}, making it a promising target for axion searches. SGR J1745–2900 is in fact frequently adopted as a benchmark for developing projected sensitivities for future observations~\cite{Hook:2018iia,Battye:2019aco,Leroy:2019ghm,Witte:2021arp,Foster:2022fxn},
and various groups have attempted to place constraints using existing radio observations (see \eg~\cite{Battye:2021yue}).  
All analyses to date have  assumed that the charge distribution can be described by the GJ model, which is derived by determining the minimal co-rotating charge density needed to screen the product of $\vec{E} \cdot \vec{B}$. Magnetars, however, are expected to have magnetospheres which differ markedly from the standard pulsar population, with charge densities greatly exceeding the minimal co-rotating density. Bearing in mind that the current understanding of magnetar magnetospheres is far from complete (see \eg~\cite{turolla2015magnetars,kaspi2017magnetars} for recent reviews outlining the current understanding of magnetars), we describe below the expected properties of these systems, and use the techniques outlined in the previous section to revisit sensitivity estimates that could be achieved with existing and future data. Needless to say, the validity of these projections hinges upon a variety of rough approximations, and thus they should be treated with some skepticism; nevertheless, these projections serve to answer the important question of whether magnetars could be useful in the future to probe regions of axion parameter space which are conventionally inaccessible to other indirect axion searches.

Magnetars are very young and highly magnetized objects whose bright X-ray emission is powered magnetically, rather than rotationally (implying the energy losses exceed the spin-down power of the neutron star), see \eg, \cite{kaspi2017magnetars}. These objects are strongly variable and exhibit a broad range of phenomenon ranging from X-ray bursts, glitches and antiglitches, non-uniform spin-down, giant flares, and  even fast radio bursts (see, \eg, \cite{Bochenek:2020zxn} for a recent association between fast radio bursts and a nearby magnetar).

It is believed that the variable magnetar activity is driven by the evolution of the ultra-strong magnetic field -- a shifting in the structure of the magnetic field internal to the neutron star deforms the crust, inducing a strong shearing of the crust which subsequently drives electric currents into the magnetosphere. The net result is a twisted, nearly force-free, magnetosphere threaded by strong electric currents $\vec{j} \sim \nabla \times \vec{B}$. Phenomena like flaring can then be explained, \eg, via the presence of instabilities that appear as the magnetic twist exceeds a critical threshold (see \eg, \cite{thompson2002electrodynamics,turolla2015magnetars,kaspi2017magnetars,Thompson:2020hwt}).

\begin{figure*}
	\centering
	\includegraphics[width=0.30\textwidth]{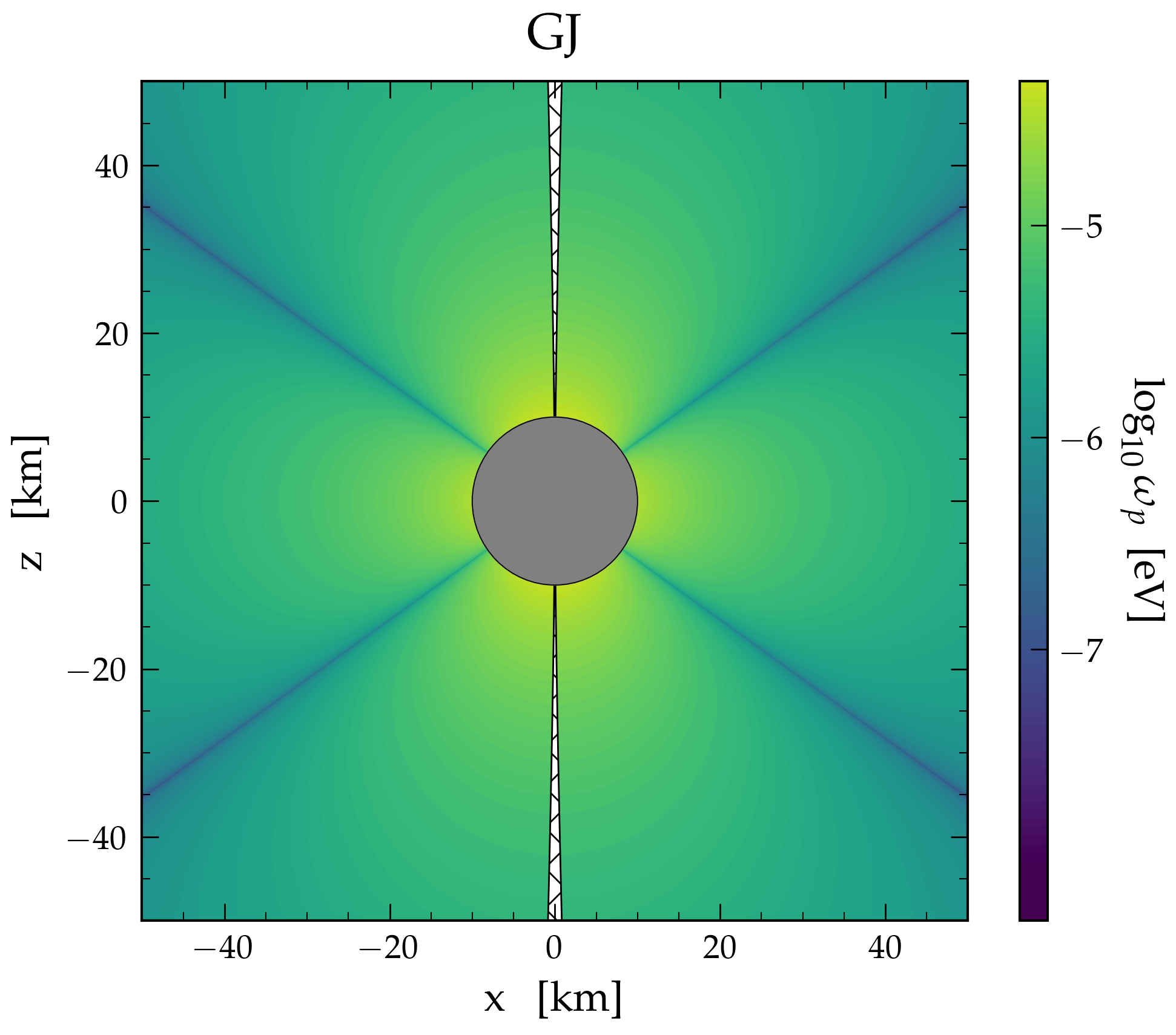}
	\includegraphics[width=0.30\textwidth]{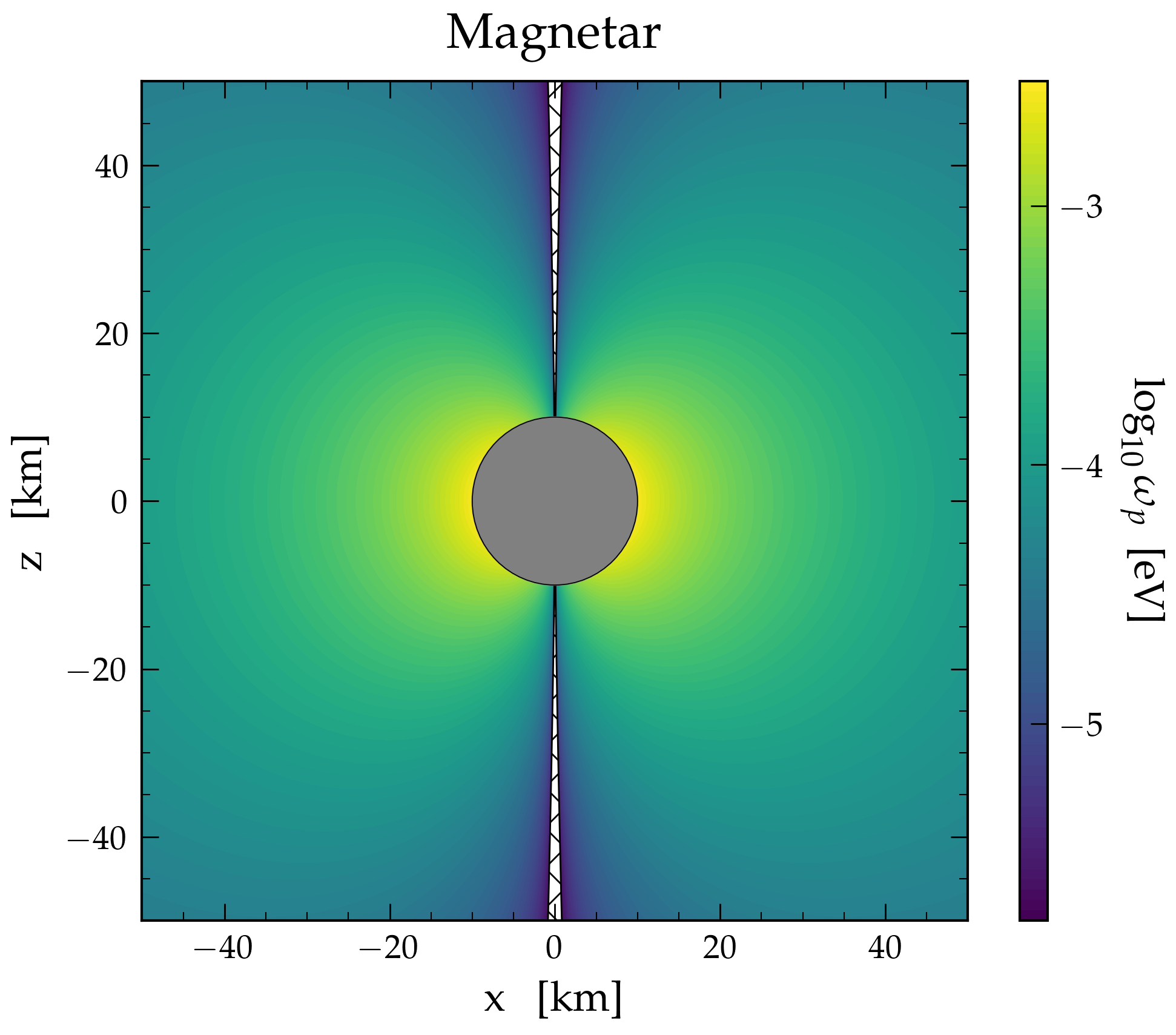}
	\includegraphics[width=0.30\textwidth]{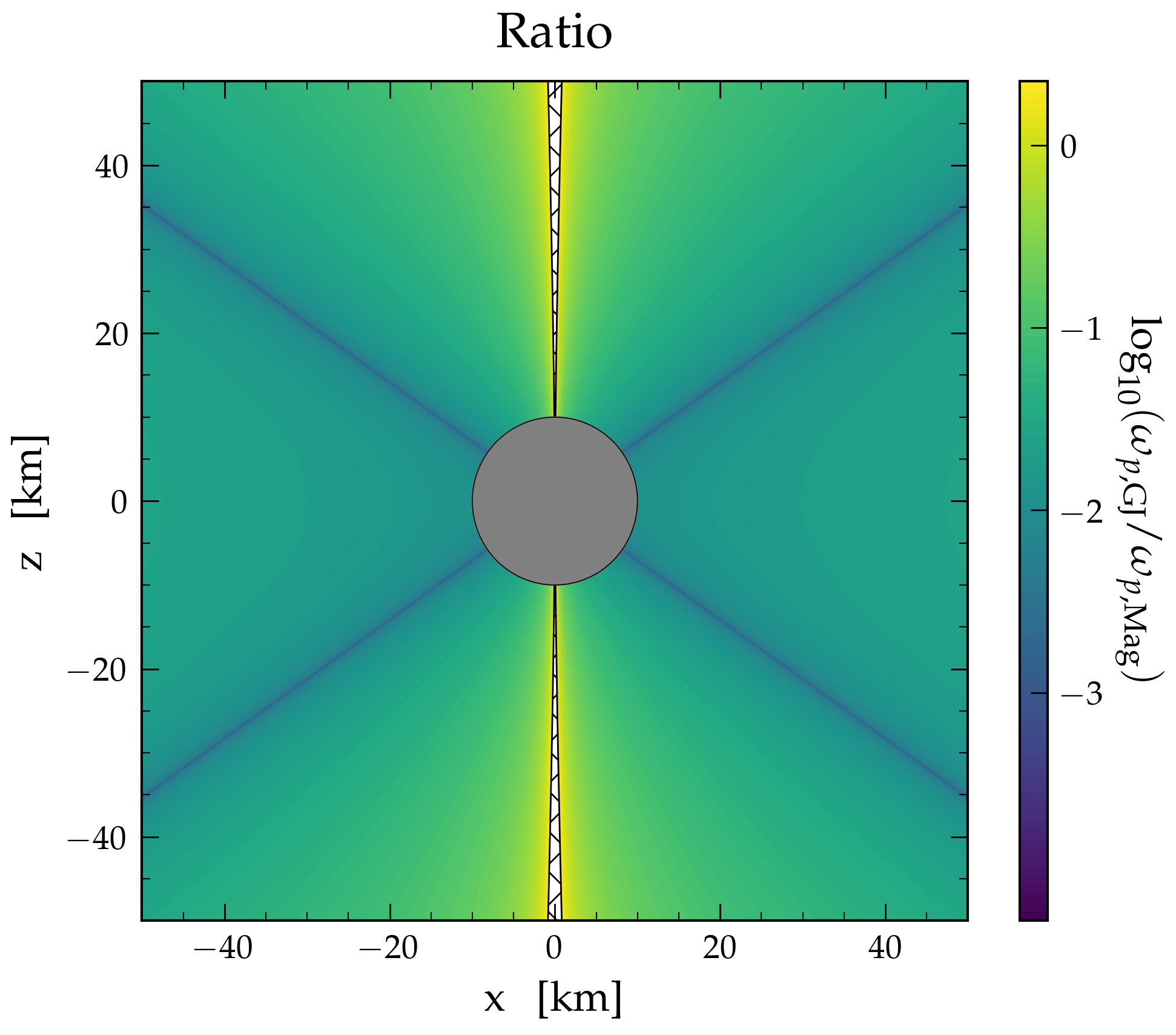}
	\caption{\textbf{Magnetar Plasma Distributions.} Left: Log-10 of the plasma frequency in the GJ model for an aligned rotator with $B_0 = 1.6\times10^{14}$ G and $P=3.76$ s. The vacuum regions defining charge separation in the GJ model can clearly be seen extending across the diagonals. Center: Log-10 of the plasma frequency as computed using Eq.~\eqref{eq:neMag} for the same parameters. Right: Log-10 of the ratio of the plasma frequency in the preceding models. In all cases the open field lines have been excised (shown with white hatched region) as the charge densities are expected to differ notably from the minimal force-free values in this region.
	}
	\label{fig:PlasDensity}
\end{figure*}

In the context of axions, there are two important distinctions between the magnetospheres of magnetars and those of standard pulsars. First, the assumption of a purely dipolar field is broken -- the twisted magnetosphere induces a complex topology to the magnetic structure, and the magnetic field strength may easily exceed the inferred dipolar value~\cite{stella2005gravitational,pons2011magnetars,rea2012new,makishima2014possible,igoshev2021strong}.  The second important distinction comes from the fact that the minimum charge density flowing from the twisted field configuration greatly exceeds the minimum co-rotational charge density identified in the GJ Model. This can be seen by noting that the force-free condition, $\vec{E} \cdot \vec{B} = 0$, implies a minimal charge density near the neutron star given by~\cite{thompson2002electrodynamics}
\begin{eqnarray}
	\rho = \nabla \cdot \vec{E} =  \Omega \cdot \left[-2\vec{B} + \vec{r} \times (\nabla \times \vec{B}) \right] = \rho_{\rm GJ} + \rho_{\rm tws}\, ,
\end{eqnarray}
where the first term is the GJ charge density, and the second $\rho_{\rm tws}$ is the minimal charge density required to support a twisted field configuration; note that for non-twisted field configurations (as, e.g., is the case with a standard dipolar field) $\nabla \times \vec{B} = 0$, and thus $\rho \sim \rho_{\rm GJ}$ (at least in the closed zone near the neutron star, where the force free condition is expected to be satisfied). The twisted field configuration expected to arise in the magnetospheres of magnetars is supported and stabilized by the presence of a strong electromagnetic current $|j| \sim \nabla \times \vec{B} \gg \rho_{\rm GJ}$, which is sourced from $e^\pm$ pair production processes near the neutron star. The electron/positron number density $n_{e^\pm}$ can be derived by writing the local charge and current density as
\begin{eqnarray}
    \rho &=& e (n_+ - n_-) \\
    |\vec{j}| &=& e (n_+ \, v_+ - n_- \, v_-) \, ,
\end{eqnarray}
where $n_\pm$ and $v_\pm$ are the number density and velocity of the $e^\pm$, and then directly solving for $n_+ + n_-$; for a semi-relativistic plasma~\cite{Thompson:2020hwt} with $|j| \gg \rho$ this quantity is roughly bounded to be $n_{\pm} \gtrsim 2 |\vec{j}| / e$~\cite{thompson2002electrodynamics,Beloborodov:2006qh}.

Ref.~\cite{thompson2002electrodynamics} investigated the charge distribution in the context of a globally twisted dipole configuration, showing  this leads to a characteristic current on the order of (see also~\cite{Beloborodov:2006qh})
\begin{equation}\label{eq:jtwist}
\vec{j}(r, \theta) = \nabla \times \vec{B} \simeq \frac{\sin^2\theta \, \psi}{r} \, \vec{B} \, .
\end{equation}
 Introducing a charge multiplicity factor $\lambda$ (defined with respect to the minimal charge density $\sim 2 |\vec{j}| / e$) and assuming the magnetic field strength is purely a function of radius, decaying as a dipolar magnetic field  $|B| \sim B_0 \, (r_{\rm NS} / r)^3$, one finds
\begin{eqnarray}\label{eq:neMag}
n_e &  \sim & \lambda \frac{\psi}{e \, r} \sin^2\theta \, B_0 \, \left(\frac{r_{\rm NS}}{r} \right)^3  \\
& \sim & \frac{7 \times 10^{15}}{\rm cm^{3}}\, \lambda \, \sin^2\theta \,  \Big(\frac{\psi}{0.2} \Big) \, \Big(\frac{B_0}{2 \times 10^{14} \, {\rm G}} \Big) \, \Big(\frac{r_{\rm NS}}{r} \Big)^4 \nonumber \, .
\end{eqnarray}
Near the neutron star surface, the charge density implied by Eq.~\eqref{eq:neMag} can exceed the GJ value (see Eq.~\eqref{eq:GJDensity})
\begin{eqnarray}
	n_e^{\rm GJ} \sim \frac{2 \times 10^{12}}{{\rm cm}^3} \Big(\frac{3.76}{P} \Big)\Big(\frac{B_0}{2 \times 10^{14} \, {\rm G}} \Big)\Big(\frac{r_{\rm NS}}{r} \Big)^3
\end{eqnarray}
by a factor $n_e / n_{e}^{\rm GJ} \sim 10^3 \times \lambda$, which is in agreement with inferences of the charge densities obtained from observations of the resonant cyclotron absorption of X-rays~\cite{Rea:2008zs}. Notice that Eq.~\eqref{eq:neMag} implies the magnetospheres of magnetars may support resonances for  axions with masses near, and above, $10^{-2} \, {\rm eV}$.

One of the fundamental questions sitting at the forefront of the field for many years is how such strong currents are sustained in magnetar magnetospheres. Recent work on the electrodynamics in super-QED field strengths has shown that the hard X-ray spectrum extending to energies $E \gtrsim 10$ keV observed in many magnetars can arise from a highly collisional semi-relativistic plasma with a characteristic density $10-20$ times larger than the minimal current density given in Eq.~\eqref{eq:jtwist}. These enhanced densities are sustained via a combination of ohmic heating and pair creation, and may be necessary in order to explain a number of observed phenomena including rapid X-ray brightening, concentrated thermal hotspots, and thermal X-ray emission~\cite{Thompson:2020hwt}.

The goal of this section is not to provide an accurate description of axion conversion in magnetar magnetospheres, but merely to point out the extent to which the axion searches performed in, \eg, Ref.~\cite{Battye:2021yue}, are modified when more realistic assumptions are adopted. In this vein, we take four fiducial models, which, in spite of their simplicity, are expected to give some rough estimation of the sensitivity that radio experiments could have to axion-photon conversion in these systems. These models consider two distinct values of magnetic fields, one with the minimal value $B_0 = 1.6 \times 10^{14} $ G (corresponding to the pure dipolar magnetic field) and the other with $B_0 = 4 \times 10^{14} $ G (while this number is somewhat \emph{ad hoc}, intended to show the impact of a moderate magnetic enhancement although an order one enhancement, we note that it could be seen as roughly consistent with the twist factor inferred in~\cite{Zelati:2015vya}), and two distinct values of charge multiplicity parameter $\lambda$ (specifically, we take a uniform value of $\lambda = 1$, and $\lambda = 20$); note that the $\lambda = 1$ is intended to represent a minimal lower bound, with $\lambda \sim \mathcal{O}(10)$ being closer to the value predicted by~\cite{Thompson:2020hwt}. We therefore consider four magnetar models, M1: $\lambda = 1, B = 1.6\times10^{14}$ G, M2: $\lambda = 20, B = 1.6\times10^{14}$ G, M3: $\lambda = 1, B = 4\times10^{14}$ G, M4: $\lambda = 20, B = 4\times10^{14}$ G.

In order to simplify the analysis, we treat the magnetic field as being purely dipolar (despite this assumption being inconsistent with the adopted charge density magnetic field structures); in general, the magnetic field structure should be obtained by solving the Grad-Shafranov equation (see \eg~\cite{beloborodov2009untwisting}), however including such an effect is beyond the scope of this work. We note, however, that there are three effects which are expected to arise as one includes the geometric effects appearing in twisted configurations. First, $\mathcal{O}(1)$ angular factors shift the photon production efficiency, and therefore induce comparable shifts in the photon anisotropy (as compared with a dipolar field configuration). Next, the radial dependence of twisted magnetic fields is modified with respect to the dipolar case (falling as $r^{-2 + p}$, with $p < 1$ for a twisted  field and $p = 1$ for the dipolar field)~\cite{thompson2002electrodynamics,Lyutikov:2021xsz}. Finally, and most importantly, the optical depth for highly twisted fields can be increased. For the moment, we estimate the optical depth using the dipolar configuration but caution that a more careful assessment of this effect may be needed in the future.

In order to be conservative, we choose to remove axion-photon conversion in open magnetic field lines, since active pair production and current flows in these regions are expected to require a more sophisticated level of modeling. These excised regions correspond to the white hatching in Fig.~\ref{fig:PlasDensity}. Using Eq.~\eqref{eq:neMag} we see that field lines are characterized by the curves $r / \sin^2\theta = L$, where $L$ gives the maximal radial distance of the field line from the neutron star. Open field lines are those which extend beyond the light-cylinder at $r_{\rm LC} = \Omega^{-1}_{\rm NS}$, i.e., they satisfy $L \geq \Omega_{\rm NS}^{-1}$. When considering axion-photon conversion, we do not include photon production occurring on open magnetic field lines.


As a word of caution, dense return currents running along the closed field line could produce secondary effects not included by this `cutting' procedure, such as the redirection and funneling photons near the conversion surface. These effects are not included here, but will be investigated in future work.

Resonant cyclotron absorption, occurring when the frequency of the radiation matches the cyclotron frequency $\omega = \omega_c$, can be increasingly important for low-frequency radiation emitted near highly magnetized objects (Ref.~\cite{Witte:2021arp} had shown using the GJ model that the optical depth can be $\mathcal{O}(1)$ for the Galactic Center magnetar). Assuming the cyclotron resonance occurs at large distances from the magnetar where the trajectories can be approximated as radial, the optical depth is roughly given by~\cite{Witte:2021arp}
\begin{equation}
\tau \sim \frac{\pi}{3} \left(\frac{\omega_p^2}{\omega} \right) \, r \ ,
\end{equation}
where it is understood that all quantities are evaluated at the point of resonance.

\begin{figure*}
	\centering
	\includegraphics[width=0.46\textwidth]{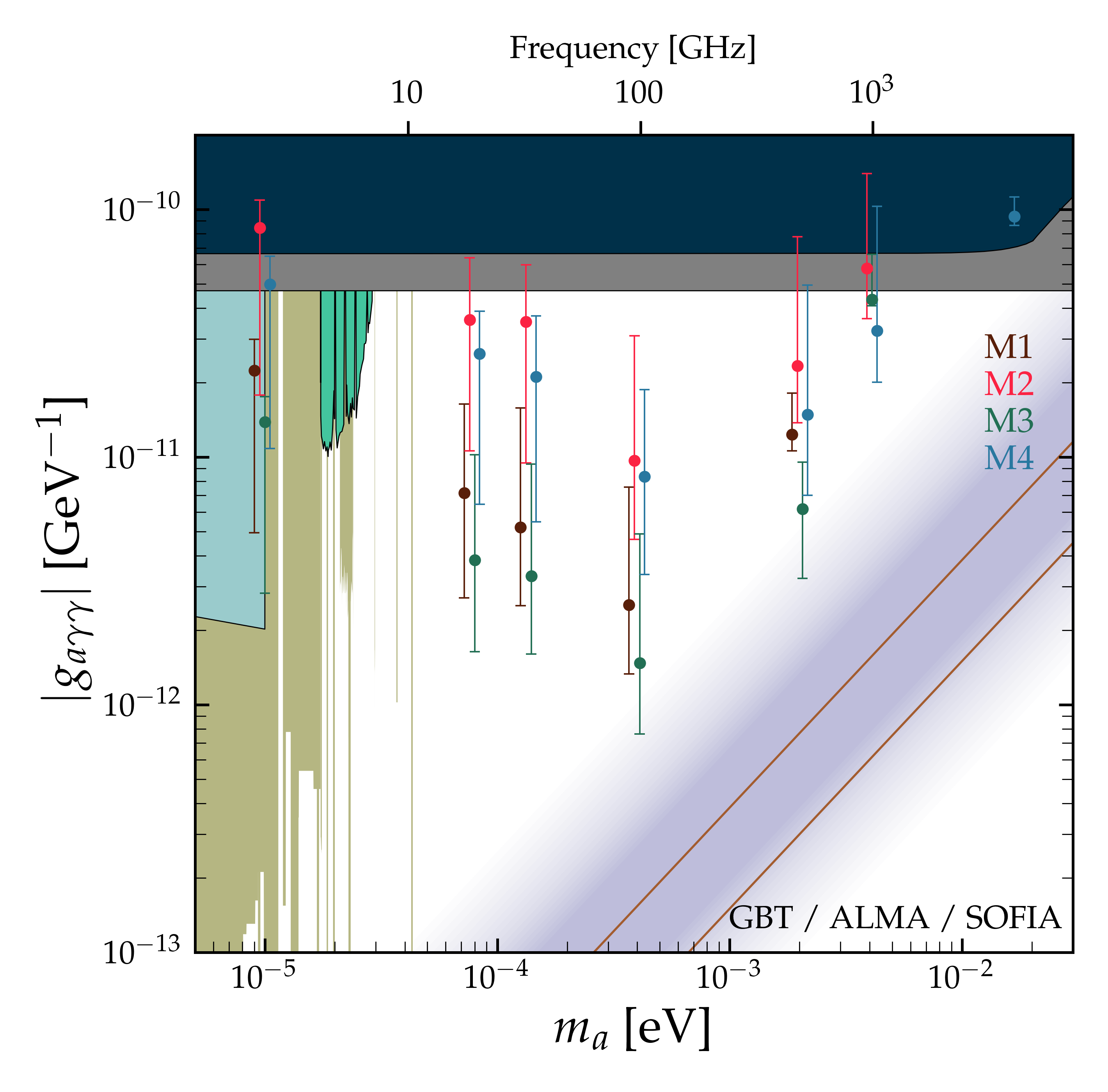}
 \includegraphics[width=0.46\textwidth]{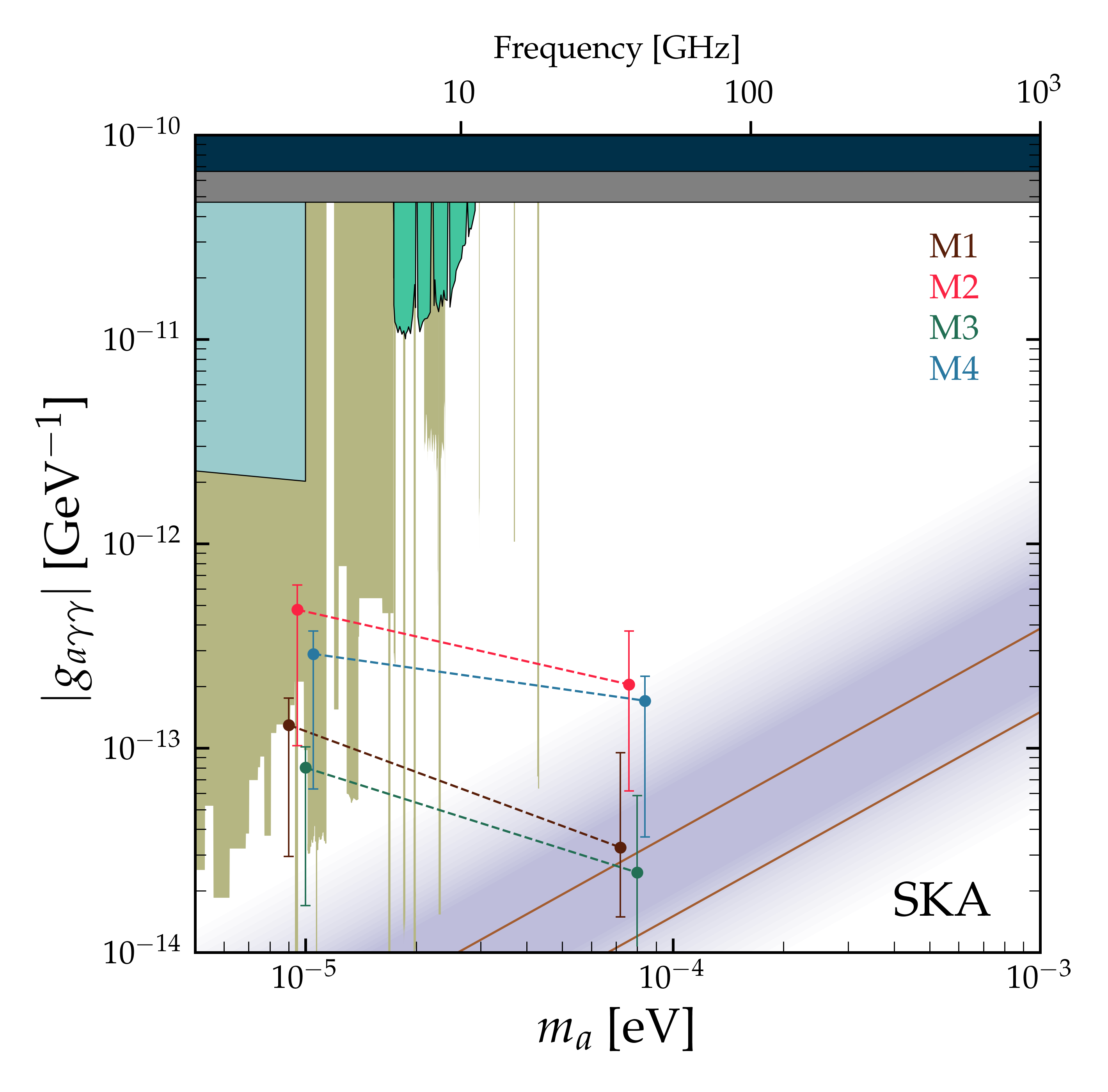}
	\caption{\textbf{Axion Searches with Galactic Center Magnetar.} Projected sensitivity to the Galactic Center magnetar SGR 1745-2900 using four distinct models based on the magnetar charge distribution given in Eq.~\eqref{eq:neMag}; model M1 is defined taking $\lambda = 1, B = 1.6\times10^{14}$ G, model M2 for $\lambda = 20, B = 1.6\times10^{14}$ G, model M3 with $\lambda = 1, B = 4\times10^{14}$ G, model M4 is defined with $\lambda = 20, B = 4\times10^{14}$ G. Results are shown assuming a distance of $d=8.3$ kpc, an NFW profile (with $r_s = 20$ kpc, and $\rho_\oplus = 0.346 \, {\rm GeV/cm^3}$), $M_{\rm NS} = 1.4 \, M_\odot$, $r_{\rm NS} = 12$km, and $\theta_m = 0$.  Left panel corresponds to sensitivity that could be achieved using current telescopes (namely a combination of GBT, ALMA, and SOFIA), while the right panel includes projected sensitivity from SKA. The vertical bars on each point reflect the $1\sigma$ variation in the inferred limits which are obtained by randomly sampling the orientation of Earth as defined with respect to the rotational axis. Shown for reference are current the QCD axion band (purple)~\cite{DiLuzio:2020wdo}, and constraints from globular clusters (gray)~\cite{Dolan:2022kul}, CAST (dark blue)~\cite{Anastassopoulos2017}, axion haloscopes (gold)~\cite{Sikivie1983,DePanfilis:1987,Hagmann:1990,Hagmann:1998cb,Asztalos:2001tf,Asztalos:2009yp,Du:2018uak,Braine2020,Bradley2003,Bradley2004,Shokair2014,HAYSTAC,Zhong2018,Backes_2021,mcallister2017organ,QUAX:2020adt,Choi_2021,Alvarez_Melcon_2021}, pulsar polar cap cascades (light blue) ~\cite{Noordhuis:2022ljw}, and GBT observations of the Galactic Center (green)~\cite{Foster:2022fxn}.
	\label{fig:maglim}}
\end{figure*}

The characteristic charge densities spanned by Eq.~\eqref{eq:neMag} suggest that magnetars will be efficiently producing radiation across frequencies from $\mathcal{O}(1)$ GHz - $\mathcal{O}(5)$ THz, and thus we use a combination of sub-mm telescopes to develop our sensitivity projections. In particular, we adopt projections for current telescopes based on observations by the Green Bank Telescope (GBT), the Atacama Large Millimeter Array (ALMA), and the Stratospheric Observatory for Infrared Astronomy (SOFIA), which have broad bandwidth coverage over the $\mathcal{O}(10)$GHz-THz regime. In addition, we include a separate set of projections for the Square Kilometer Array (SKA), which will cover frequencies from 50 MHz to 24 GHz. We compute the radio spectrum at seven fixed axion masses, corresponding to observing frequencies of $\sim$ 2.4, 20, 35, 100 GHz, 500 GHz, 1 THz, and 4 THz. SKA observations will cover the lowest two frequency bins, and assuming a system equivalent flux density SEFD = 0.098 Jy, a bandwidth of $10^{-5} \times m_a$, and an observing time of $10$ hours, SKA will have sensitivity (at the 95$\%$ confidence level) to radio lines at each observing frequency of $5$ and $2 \, \mu$Jy~\cite{SKA}. We use the GBT telescope to establish sensitivity in the lowest three frequencies  -- here, we adopt a sensitivity (across all frequencies) consistent with the  quoted $95\%$ confidence upper limit used in the analysis of~\cite{Battye:2021yue}, $S_{\rm lim} \sim 0.3$ mJy (computed using a bin width of $\delta f \sim 28$ MHz). At 100 \footnote{GBT can also observe at 100 GHz -- using the online sensitivity calculator for GBT, we find a couple hours of observation tends to give comparable sensitivity estimates.} and 500 GHz, we adopt a sensitivity for ALMA consistent with the quoted capability for $60$ seconds of observations and a line width of \footnote{The line studied here is expected to be slightly wider than this level, however the sensitivity scales weakly with bandwidth, and can easily be compensated for using additional observing time.} $\delta f/ f \sim 10^{-6}$,
which corresponds to $S_{\rm lim} \sim$ 5 mJy and 25 mJy for 100 and 500 GHz respectively~\cite{wootten2003atacama}. Note that GC magnetar has been observed up to frequencies of a few hundred GHz, with the observed flux density sitting below the 10 mJy level~\cite{torne2016detection}. The sensitivity of the two highest frequency bins is set to 100 mJy, which is roughly the $4\sigma$ line sensitivity for 900 seconds of observation estimated in~\cite{gehrz2009new}. In general, one would either marginalize over the unknown parameters, or take the value which give the most conservative constraints \cite{Battye:2021yue,Battye:2023oac} -- since our goal, however, is only to provide an indicative idea of rough sensitivity, we simply fix $\theta_m = 0$, $d = 8.3$ kpc, $M_{\rm NS} = 1.4 \, M_\odot$, and $r_{\rm NS} = 12$ km. Note that since we do not vary $\theta_m$, which plays an important role in determining the line width, we make the simplifying assumption that the entirety of the signal is contained within a single frequency bin -- this bin is assumed to have a value of $10^{-5} \times m_a$ for all telescopes except GBT, where we take the bin width used in the observations of Ref.~\cite{Battye:2021yue}.

The projected sensitivity to the axion-photon coupling in each of our four magnetar models M1-M4 motivated above, is shown in Fig.~\ref{fig:maglim}. In this figure we show the flux predicted at a typical viewing angle, defined by generating $10^3$ samples and selecting the median value, and the $\pm 1\sigma$ variations about the median value. Fig.~\ref{fig:maglim} illustrates that SGR 1745-2900 may produce observable radio emission up to $\sim 4$THz, however only if there exists a sizable non-dipolar contribution to the magnetic field and the charge density exceeds the minimal expected value by an order 10 value -- nevertheless, emission up to $\sim 500$ GHz is still expected across all models, potentially allowing parts of the QCD axion band to be explored using existing instruments. These sensitivities should still be interpreted with caution, as systematic uncertainties have not been folded in, and the impact of non-dipolar field modeling has not yet been explored.

\section{Conclusions}\label{sec:conclusions}

In this work, we have constructed a generalized ray tracing framework capable of analyzing radio signals sourced from resonant axion-photon mixing in astrophysical plasmas, focusing in particular on the treatment of these interactions in highly magnetized plasmas and curved spacetime. We have explicitly shown how these calculations can be self-consistently embedded using either a `forward ray tracing' approach (in which one samples from the photon phase-space at production, propagates the photons to far distances, and reconstructs observables from the final photon distribution) or a `backward ray tracing' approach (the more conventional ray tracing approach, in which one propagates rays from an observing plane far away from the source to the point of production); while these approaches use different methodology, we have shown using detailed phase-space arguments (Sec.~\ref{sec:raytrace}-\ref{sec:AxionsPlasma}) that these must yield identical results.
We then demonstrated this spectacular agreement explicitly through extensive numerical comparison of the two codes. Note that this is a highly non-trivial result, as even small deviations in the definitions of fundamental constants or accumulated errors in the ODE solvers can generate sizable effects.

Previous work on ray tracing in astrophysical axion searches have included only a subset of the effects studied here, focusing either on the propagation of photons through a magnetized plasma in flat space~\cite{Witte:2021arp} (using forward propagation) or through an isotropic plasma in curved space~\cite{Battye:2021xvt} (using backward propagation). This work unites these frameworks and allows for a thorough investigation of each of the assumptions adopted in the literature thus far. Our primary conclusions are as follows:
\begin{itemize}
	\item In curved space, the anisotropy of the plasma tends to squeeze the radiated power to small angles (but away from the magnetic pole). This can cause the observed power to deviate from the isotropic scenario by potentially an order of magnitude or so, depending on the axion mass and the viewing angle.
	\item For large neutron star and axion masses, gravity can induce sizable modifications to photon trajectories, and tends to isotropize the radiated flux. For small neutron star and axion masses this effect becomes negligible. Interestingly, we find that previous approaches which had included the effect of gravity in the initial conditions but not in the propagation of photons are extremely accurate, despite not being self-consistent.
	\item Varying the neutron radius within a range of values permitted by the equation of state can lead to a factor of $\sim 2$ shift in the sky-integrated power. This effect is predominantly driven by the change in the resonant surface area.
	\item The improved backward ray tracing algorithm now accounts for multiple photon production sites along photon worldlines (see Sec.~\ref{sec:multipleReflection}). Including this effect can enhance the total power by a factor of  a few (the effect being more pronounced for smaller axion masses). Crucially though, the inclusion of these effects is needed to have agreement between the two ray tracing approaches used in this paper.
\end{itemize}

In this work (and Ref.~\cite{McDonald:2023ohd}), we have gone to great lengths to fully develop kinetic theory in anisotropic media, which has additional complications relative to isotropic backgrounds. In particular,  generalizing the dispersion relation for photons in anisotropic plasmas to curved spacetime \cite{GedalinMelrose,BreuerEhlers1980,BreuerEhlers1981a,BreuerEhlers1981II}, is somewhat more involved that an isotropic plasma \cite{Rogers:2015dla,1975Hadrava}. Similarly, the geometry of the conversion surface is also more complicated (both in flat and curved spacetime), and consists of a foliation of multiple production surfaces. In turn, these correspond to more complicated  kinematic matching of axions and photons at the conversion surface.  

Related to this discussion of anisotropic media, we have paid particular attention to including curved spacetime effects in our routines, which are required to self-consistently incorporate gravity across the full range of physical effects. For self-consistency, gravity should also be incorporated into the conversion probability itself, such that when integrating over phase-space (see Sec. \ref{sec:AxionsPlasma}) the resulting power is convergent. In this work, and Ref.~\cite{McDonald:2023ohd}, we have shown that in flat space, both the isotropic and anisotropic conversion probability gives finite results. In Sec.~\ref{sec:curvedS}, we also offered arguments for generalising the conversion probability in isotropic plasmas to curved spacetime, showing that this generalized form of the conversion probability leads to convergent results.  One of the remaining open problems, however, remains a full generalization of the \textit{anisotropic} conversion probability to curved spacetime. Presumably the answer lies somewhere in generalising the phase-space and kinetic theory arguments of the present work using techniques laid out in Refs.~\cite{Acuna-Cardenas:2021nkj,Hohenegger:2008zk}, though we leave such derivations for future work.

With a view to observations, in Sec.~\ref{sec:NS}, we have used our newly developed ray tracing framework to revisit sensitivity estimates of radio and microwave telescopes to spectral lines emanating from Galactic Center magnetar SGR J1745-2900. Here, we provided an extensive discussion on the state-of-the-art knowledge of charge distributions in magnetars, showing that previous approximations relaying on the Goldreich-Julian charge distribution have likely underestimated the characteristic plasma frequency near the surface of the star. Using four distinct models, we show that the high plasma densities near the surface of the magnetar are capable of generating electromagnetic signatures up to the $\mathcal{O}({\rm THz})$ regime, with current and future infrastructure potentially covering significant unexplored regions of  axion parameter space. This work provides greater motivation for understanding magnetar charge densities, as such searches may provide one of the unique avenues for indirectly probing this well-motivated region of axion parameter space.

Ray tracing has proven to be an invaluable tool in astronomy and astrophysics, and in recent years has emerged an increasingly important approach in the indirect search for axions. This paper has developed the fundamentals necessary to incorporate ray tracing into astrophysical axion searches, and has for the first time investigated and quantified the validity of a variety of different assumptions adopted in previous applications and searches. While there still exist open questions which need to be addressed, such as the impact of uncertainties in the charge distribution and near-field magnetic field configuration, and how axions and photons mix in high magnetized environments, the framework developed here lays the much needed groundwork for the future development of indirect axion searches in neutron star magnetospheres.

\begin{center}
\textbf{Acknowledgements}
\end{center}

We thank Richard Battye, Avery Broderick,  Bj{\"o}rn Garbrecht, Jeremy Heyl, Pete Millington, and Oilivier Sarbach for useful discussions. J.I.M. is supported by an FSR Fellowship and thanks Utkarsh Bhura for assistance in accessing data on-site during the 2023 cyber incident at Manchester University.  SJW acknowledges support from a Royal Society University Research Fellowship (URF-R1-231065), the program
Ram\'{o}n y Cajal (RYC2021-030893-I) of the Spanish
Ministry of Science and Innovation, and through the
European Research Council (ERC) under the European Union's Horizon 2020 research and innovation
programme (Grant agreement No. 864035 Undark)
and the Netherlands eScience Center, grant number
ETEC.2019.018.). This article/publication is based upon work from COST Action COSMIC WISPers CA21106, supported by COST (European Cooperation in Science and Technology).

\appendix

\bibliographystyle{apsrev4-1}
\bibliography{ref.bib}

\appendix

\section{Anisotropic Conversion in Curved Space - an Open Problem}\label{sec:APPCurved}

\begin{figure*}
\includegraphics[width=0.49\textwidth]{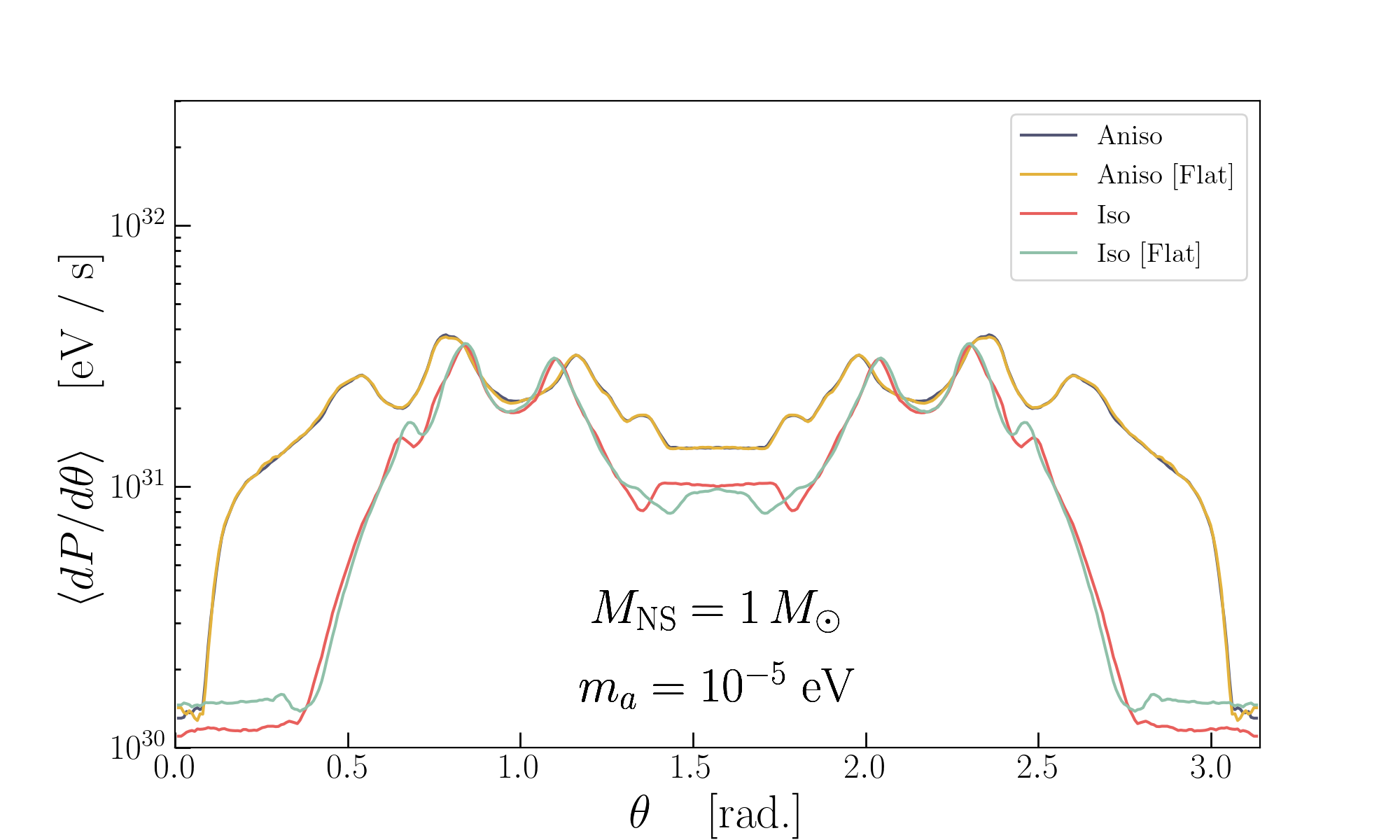}
\includegraphics[width=0.49\textwidth]{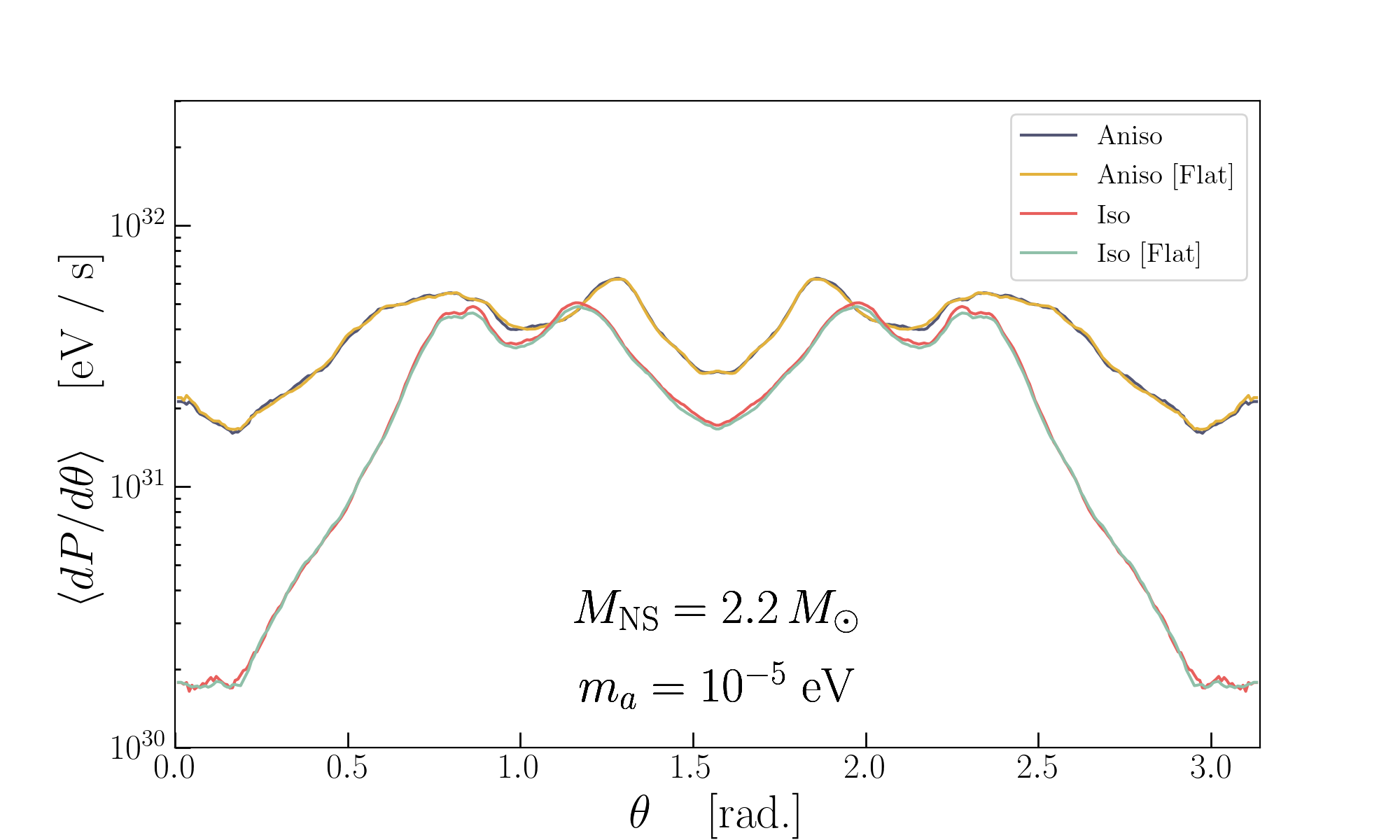}
\caption{\label{fig:Gravity} \textbf{Artificial Gravity.} Result of adopting initial conditions consistent with photon production from axions near a neutron star with mass $M_{\rm NS } = 1 \, M_\odot$ (left) and $2.2 \, M_\odot$ (right), but setting $M_{\rm NS} = 0$ in the ray tracing procedure.  }
\end{figure*}

At this point, a few remarks are in order about the present difficulties in generalising the results of this subsection to curved space time. Firstly, following results of Ref.~\cite{McDonald:2023ohd}, one could conjecture that the width of the resonance can be inferred from the governing Boltzmann equation projected along photon-wordlines, which gives
\begin{align}\label{eq:BoltzmannExplicit}
  &\frac{d f_\gamma(x(\lambda),k(\lambda)}{d \lambda}   \nonumber \\
  & = g_{a \gamma \gamma}^2 \big| k \cdot \tilde{F}_{\rm ext} \cdot \varepsilon \big|^2 2 \pi \delta \left(g^{\mu \nu}(\lambda) k_\mu(\lambda) k_\nu(\lambda) - m_a^2\right) f_a \, .
\end{align}
Clearly, in an isotropic medium, the argument of the delta function just becomes $\omega_{\rm p}^2 - m_a^2$, and we recover the result above when performing the integration over $\lambda$, which gives terms in the denominator proportional to $d \omega_p^2/d\lambda \propto k \cdot \partial_\mu (\omega_p^2)$. For a more general dispersion relation, however, the integration of the delta-function yields
\begin{eqnarray}\label{eq:CurvedfGamma1}
	f_\gamma = \frac{\pi   g_{a \gamma \gamma}^2 \big| k \cdot \tilde{F}_{\rm ext} \cdot \varepsilon \big|^2}{\left| k \cdot \mathcal{D} \mathcal{H} \right|} f_a
\end{eqnarray}
where $\mathcal{H}$ is the photon-Hamiltonian defined in \cite{McDonald:2023ohd} and
\begin{equation}
	\mathcal{D}_\mu = \frac{\partial}{\partial x^\mu}   + \Gamma^\sigma_{\mu \rho} k_\sigma \frac{\partial}{\partial k_\rho},
\end{equation}
is the generalized covariant-derivative appearing in \cite{Acuna-Cardenas:2021nkj,Hohenegger:2008zk}. To arrive at Eq.~\eqref{eq:CurvedfGamma1} we used Hamiltonians equations (Eq.~\eqref{eq:Hamilton}) for $x^\mu(\lambda)$ and $k_\mu(\lambda)$ to express their $\lambda$ derivatives in terms of derivatives of $\mathcal{H}$. We also assumed a Levi-Civita connection in which $g_{\mu \nu}$ is covariantly constant, so that $\nabla_\sigma g_{\mu \nu} = 0 $, allowing us to re-express partial derivatives of $\partial_\sigma g^{\mu \nu}$ in terms of Cristoffel symbols $\Gamma^\sigma_{\mu \rho}$. The following ansatz would then lead to a conversion probability
\begin{eqnarray}\label{eq:CurvedfGamma2}
	P_{a \gamma} = \frac{ \pi   g_{a \gamma \gamma}^2 \big| k \cdot \tilde{F}_{\rm ext} \cdot \varepsilon \big|^2}{\left| k \cdot \mathcal{D} \mathcal{H} \right|} .
\end{eqnarray}
Clearly a divergence occurs when $k$ is perpendicular to $\mathcal{D}\mathcal{H}$. One might therefore hope that this divergence is regulated by the phase-space measure appearing in Eq.~\eqref{eq:ForwardTracingEquation1}; however, the key difficulty is generalising Eq.~\eqref{eq:ForwardTracingEquation1} to curved space is to generalize the step of Eq.~\eqref{eq:IntegralTrick}, which proves difficult since one picks up additional derivatives of the metric in the term $g^{\mu \nu} k_\mu k_\nu$, which at face value do not lead to appropriate cancellations with the phase space measure. The authors suspect that there may exist a subtlety in the generalization of this formula to curved space, however this is highly non-trivial and thus it will become the subject of future work.

\begin{figure}[t!]
\includegraphics[width=0.49\textwidth]{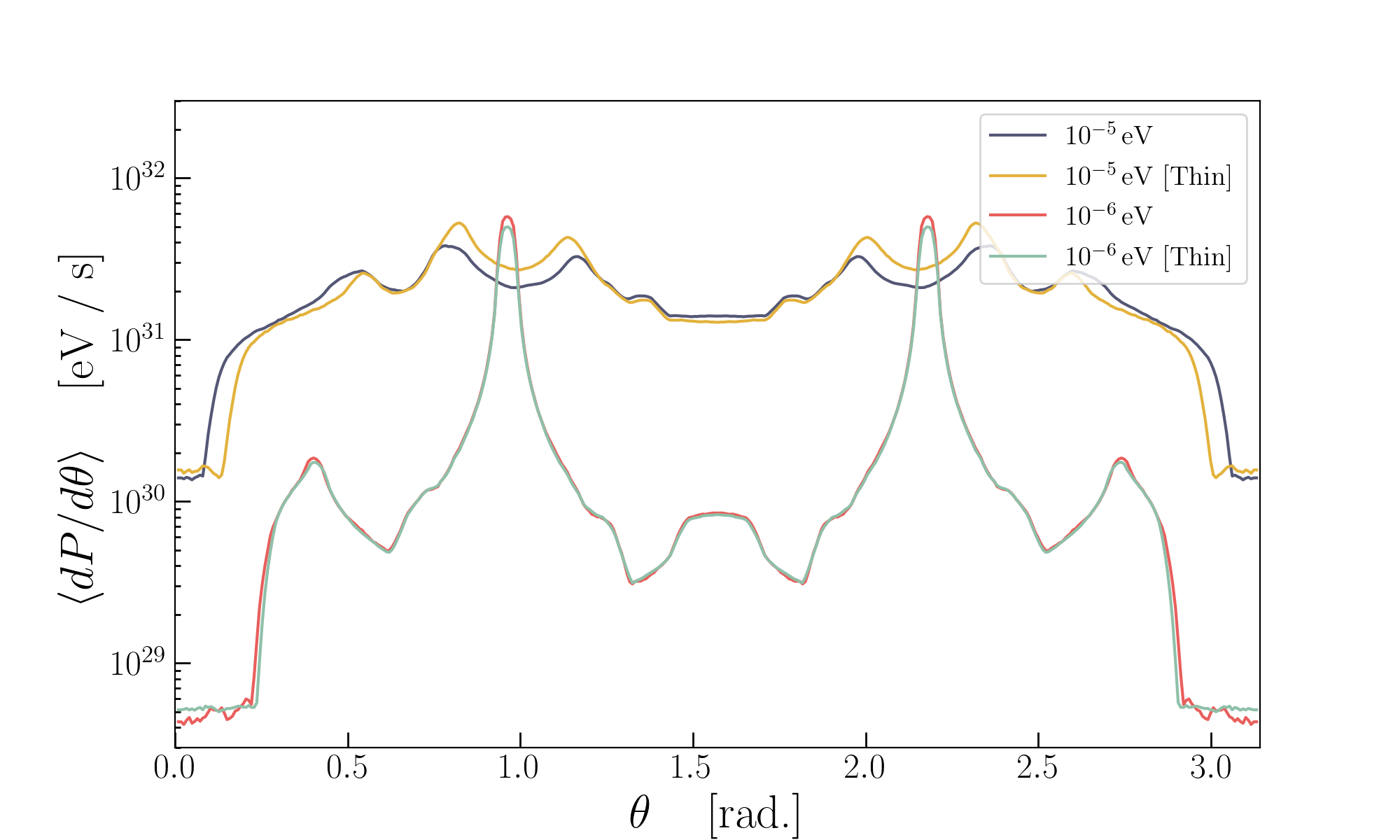}
\caption{\label{fig:Thin} \textbf{Kinematics \& Conversion Surface Geometry.} Same as left panel of Fig.~\ref{fig:flatiso}, but illustrating the impact of incorrectly imposing a resonant condition of $m_a \simeq \omega_p$, rather than $k_\mu^a = k_\mu^\gamma$.}
\end{figure}




\section{Comparison with Previous Approximation Schemes}\label{sec:APPcomparison}
In the following, we attempt to make connections with simplifying approximations adopted in previous work~\cite{Witte:2021arp}. Despite the fact that these simplified approaches are not fully self-consistent, we show the net effect on the differential power is rather minimal.

The first simplifying approximation, applied in the context of forward ray tracing~\cite{Witte:2021arp,Foster:2020pgt}, is that gravity can be embedded in the initial conditions of the photon, but neglected in the propagation. That is to say, the initial energy and momentum of the photon are consistent with being produced within a gravitational potential sourced a neutron star of mass $M_{\rm NS}$, but $M_{\rm NS}$ is set to zero in Hamilton's equations (Eq.~\ref{eq:Hamilton}). The results of performing this `flat' analysis are shown in Fig.~\ref{fig:Gravity} for two choices of neutron star mass, and both an anisotropic and isotropic dispersion relation. Here, we see that the impact of the flat approximation is typically extremely small. Importantly, this approximation scheme is not equivalent to taking $M_{\rm NS} \rightarrow 0$ (since the initial conditions for the photon momentum here are fixed in both the curved and flat space analysis), and thus we do not expect the results to mimic those of Fig.~\ref{fig:MassNS_Renorm}.

The second approximation, which was adopted in \cite{Witte:2021arp,Foster:2020pgt,Witte:2022cjj}, assumes the resonance occurs along a single two-dimensional surface appearing at $m_a \simeq \omega_p$. As mentioned in the main text (see Sec.~\ref{Sec:curvedSpacetimem}), resonances occur across a foilation of surfaces, defined by $k_{\mu}^a = k_\mu^\gamma$, which can extend to slightly larger and smaller radii than what one would infer by applying the former approximation. In Fig.~\ref{fig:Thin}, we show the relative importance of including, or neglecting, the proper kinematic matching condition. Note that in the latter case, in order to ensure the photon in on-shell, we set initial conditions of photons by taking $\omega_\gamma = \omega_a$ and the unit 3-momentum vectors to satisfy $\hat{k}_a = \hat{k}_\gamma$. The normalization $\left|\textbf{k}\right|$ of the photon 3-momentum is then inferred from Eq.~\eqref{eq:AnisotrpoicDisp}. The radial width of the foliation of surfaces defined by the appropriate resonance condition $k_{\mu}^a = k_\mu^\gamma$ tends to be below the 10$\%$ level, and thus the effect is not expected to be large; this expectation is confirmed in Fig.~\ref{fig:Thin}, which shows that the differential power is only slightly modified for a narrow region of viewing angles.

\end{document}